\documentclass[a4paper, 11pt, abstracton, titlepage, oneside]{scrartcl}

\usepackage[T1]{fontenc}
\usepackage[utf8]{inputenc}
\usepackage[onehalfspacing]{setspace}
\usepackage[headsepline]{scrlayer-scrpage}
\clearpairofpagestyles 
\usepackage{layout}
\usepackage{geometry}
\usepackage{adjustbox}
\usepackage{authblk}
\usepackage[titletoc,title]{appendix}

\usepackage[hyphens]{url}
\usepackage[colorlinks=true, allcolors=blue]{hyperref}
\usepackage{color}

\usepackage{amsmath,amssymb,amsfonts,amsthm, mathtools, bm}
\usepackage{thmtools, thm-restate}
\usepackage{nicefrac}
\usepackage{array}

\usepackage[normal]{threeparttable}
\usepackage{threeparttablex}
\usepackage{tabularx}
\usepackage{longtable}
\usepackage{booktabs}
\usepackage{colortbl}
\usepackage{multirow}
\usepackage{makecell}

\usepackage[ruled, vlined, linesnumbered]{algorithm2e} 
\SetKw{Continue}{continue}

\usepackage[acronym]{glossaries}

\usepackage{subcaption}
\usepackage{floatrow}
\usepackage{graphicx}
\usepackage{placeins}

\usepackage{tikz}
\usetikzlibrary{external}
\usetikzlibrary{arrows.meta}
\usetikzlibrary{math}
\usetikzlibrary{shapes}
\usepackage{pgfplots}
\usepgfplotslibrary{groupplots}
\usepgfplotslibrary{dateplot}
\pgfplotsset{compat=1.18} 
\usepackage{tikzscale}

\usepackage{enumerate}
\usepackage{multicol}
\usepackage[author=Patrick]{pdfcomment}
\usepackage{xparse}
\usepackage{twoopt}
\usepackage{etoolbox}

\usepackage{eurosym}
\usepackage{ellipsis}
\usepackage{natbib}
\usepackage{paralist}
\usepackage{ulem}

\AtBeginDocument{
	\newgeometry{
		textheight=9in,
		textwidth=6.5in,
		top=1in,
		headheight=14pt,
		headsep=25pt,
		footskip=30pt
	}
}
\newsavebox{\abstractbox}

\makeatletter
\expandafter\patchcmd\csname\string\maketitle\endcsname
{\vskip\z@\@plus3fill}
{\vskip\z@\@plus2fill\box\abstractbox\vskip\z@\@plus1fill}
{}{}
\makeatother

\addtokomafont{captionlabel}{\footnotesize\bfseries} 
\addtokomafont{caption}{\footnotesize\bfseries} 

\bibpunct[, ]{(}{)}{,}{a}{}{,}%
\DeclareMathOperator*{\argmax}{arg\,max}

\counterwithin*{equation}{section}

\newenvironment{myfont}{\fontfamily{ccr}\selectfont}{\par}
\DeclareTextFontCommand{\textmyfont}{\myfont}

\AtBeginDocument{%
	\abovedisplayskip=7.5pt plus 0pt minus 0pt
	\abovedisplayshortskip=7.5pt plus 0pt minus 0pt
	\belowdisplayskip=7.5pt plus 0pt minus 0pt
	\belowdisplayshortskip=7.5pt plus 0pt minus 0pt
}

\newcolumntype{L}[1]{>{\raggedright\let\newline\\\arraybackslash\hspace{0pt}}p{#1}}
\newcolumntype{C}[1]{>{\centering\let\newline\\\arraybackslash\hspace{0pt}}p{#1}}
\newcolumntype{R}[1]{>{\raggedleft\let\newline\\\arraybackslash\hspace{0pt}}p{#1}}
\renewcommand{\emph}[1]{\textit{#1}}

\begin{document}
















\newacronym{pnb}{P\&B}{price-and-branch}


\newacronym{rki}{RKI}{Robert Koch Institute}









\newcommand{\mCardinality}[1]{|{#1}|}
\newcommand{\mSum}[2]{\ensuremath{\sum_{#1}^{#2}}}

\newcommand{\mTimeHorizon}{\ensuremath{\mathcal{T}}}

\newcommand{\mGraph}{\mathcal{G}}
\newcommand{\mNodes}{\mathcal{N}}
\newcommand{\mNodesRouteLayer}{\mNodes_R}
\newcommand{\mNodesWaitingLayer}{\mNodes_S}
\newcommand{\mArcs}{\mathcal{A}}
\newcommand{\mArcsRouteLayer}{\mathcal{A}_R}
\newcommand{\mArcsWaiting}{\mathcal{A}_S}
\newcommand{\mArcsTransition}{\mathcal{A}_T}
\newcommand{\mArcsWalking}{\mathcal{A}_W}
\newcommand{\mArcsOfRoute}[1]{\mathcal{A}_{R}^{#1}}
\newcommand{\mDemand}[2]{d^{#1}_{#2}}
\newcommand{\mRouteCapacity}[2]{\kappa_{#1#2}}
\newcommand{\mFlow}[3]{f^{#1}_{#2#3}}
\newcommand{\mDecision}[3]{t^{#1}_{}}
\newcommand{\mRoutes}{\mathcal{R}}
\newcommand{\mUncertainty}{\xi}
\newcommand{\mChanceProbabilty}{\delta}
\newcommand{\mIncidenceFactor}{\omega}
\newcommand{\mDemandFactor}{\delta}
\newcommand{\mTravelTime}[2]{\tau_{#1#2}}

\newcommand{\mSetArcs}{\mathcal{A}}
\newcommand{\mSetVertices}[0]{\ensuremath{\mathcal{V}}}
\newcommand{\mSetPassengers}[0]{\ensuremath{\mathcal{P}}}
\newcommand{\mSetTransitRoutes}[0]{\ensuremath{\mathcal{R}}}
\newcommand{\mSetModes}[0]{\ensuremath{\mathcal{M}}}
\newcommand{\mSetPaths}[0]{\ensuremath{\mathcal{L}}}
\newcommand{\mBudget}[0]{\ensuremath{\zeta}}
\newcommand{\mAllocation}[0]{\ensuremath{z}}
\newcommand{\mArcCapacity}[0]{\ensuremath{\kappa}}

\begin{titlepage}
\centering

\vspace*{2em}

{\LARGE\bfseries
Public Transport Under Epidemic Conditions:\\[0.4em]
 Nonlinear Trade-Offs Between Risk and Accessibility\par}

\vspace{2em}

{\large
Gerhard Hiermann$^{1,2}$,
Joana Ji$^3$,
Ana Moreno$^{3}$,
Rolf Moeckel$^{3,\dagger}$,
Maximilian Schiffer$^{2,4}$\par}

\vspace{1em}

{\small
\small $^{1}$ Institute of Production and Logistics Management, Johannes Kepler University Linz, Austria \\
\small $^{2}$ TUM School of Management, Technical University of Munich, Germany \\
\small $^{3}$ School of Engineering \& Design, Technical University of Munich, Germany\\
\small $^{4}$ Munich Data Science Institute, Technical University of Munich, Germany\\
$^{\dagger}$ Deceased
\par}

\vspace{2.5em}

\begin{minipage}{0.9\textwidth}
\small
\textbf{Abstract.}
Epidemics expose critical tensions between protecting public health and maintaining essential urban mobility. Public transport systems face this dilemma most acutely: they enable access to jobs, education, and services, yet also facilitate close contact among travelers. We develop an integrated modeling framework that couples agent-based epidemic simulation (EpiSim) with an optimization-based public transport flow model under capacity constraints. Using Munich as a case study, we analyze how combinations of facility closures and transport restrictions shape epidemic outcomes and accessibility. The results reveal three key insights. First, epidemic interventions redistribute rather than simply reduce infection risks, shifting transmission to households. Second, epidemic and transport policies interact nonlinearly—moderate demand suppression can offset large capacity cuts. Third, epidemic pressures amplify temporal and spatial inequalities, disproportionately affecting peripheral and peak-hour travelers. These findings highlight that blanket restrictions are both inefficient and inequitable, calling for targeted, time- and space-differentiated measures to build epidemic-resilient and socially fair transport systems.
\end{minipage}
\begin{center}
\end{center}

\vfill
\end{titlepage}

\section{Introduction}
Epidemics pose significant challenges to modern societies, affecting not only public health but also the functioning of critical infrastructure and the stability of economic activity. Among these infrastructures, public transport systems (PT) are particularly vulnerable. They serve as vital arteries for urban mobility, ensuring access to work, education, and essential services, while simultaneously acting as potential vectors for disease transmission. During an epidemic, public transport faces a dual challenge: reducing the risk of infection among passengers while maintaining essential mobility for the population. This dilemma becomes especially pronounced when capacity restrictions are introduced to enforce social distancing. Such measures directly constrain the supply of transport services, while the epidemic itself simultaneously alters travel demand through infection rates, quarantines, and changing perceptions of risk. Understanding these complex interdependencies is crucial for designing resilient and equitable urban mobility systems.  

The COVID-19 pandemic brought these tensions into sharp focus. Remote work, school closures, and lockdowns substantially reduced ridership, but the simultaneous introduction of capacity limits created bottlenecks that disproportionately affected those dependent on public transport. These pressures were not evenly distributed: central areas often retained multiple mobility options, while peripheral neighborhoods experienced more severe accessibility losses due to reduced service levels and longer commutes. Such observations underscore the need for systematic analysis of how epidemic dynamics, demand changes, and transport capacity restrictions interact, and how these interactions shape both efficiency and equity in urban transport systems.  

This paper addresses this need by combining two complementary modeling approaches: a simulation-based model of epidemic spread and an optimization-based model of passenger flows in the public transport network under capacity constraints. The integration of these methods enables us to capture both the behavioral dynamics of epidemics and the structural limits of the transport system. While agent-based assignment models provide rich behavioral insights, optimization offers a tractable system-level benchmark of feasible flows, allowing us to evaluate how accessibility is constrained under epidemic conditions.  

We apply this framework to a case study of Munich, Germany, analyzing multiple epidemic and policy scenarios that reflect varying degrees of restrictions. Our analysis reveals several key findings. First, epidemic restrictions alter not only the scale but also the distribution of risks: once external contacts are curtailed, infections concentrate in households, making the role of transport restrictions primarily indirect, through their effects on mobility and accessibility. Second, epidemic and transport policies interact in nonlinear ways. Capacity cuts without demand suppression create severe accessibility bottlenecks, while even modest demand reductions through facility closures can alleviate pressure on the system. Third, epidemic pressures amplify existing inequalities: bottlenecks are concentrated during commuting peaks, and peripheral boroughs, long-distance commuters, and groups with less flexibility in their travel patterns face disproportionate burdens. Taken together, these results show that uniform, blanket restrictions are both inefficient and inequitable. More nuanced, epidemic-resilient policies—differentiated across time and space, and coordinated between demand- and supply-side measures—are needed to balance health protection, mobility provision, and social equity.

This study contributes to understanding how epidemics and transport systems interact by providing an integrated, city-scale modeling framework that links behavioral epidemic dynamics to structural accessibility constraints. Methodologically, the framework combines agent-based epidemic simulation with optimization-based passenger flow analysis, thereby capturing both individual contact processes and network-level capacity effects. Applied to Munich during the first wave of COVID-19, the model allows us to examine how different combinations of facility closures and public transport restrictions affect mobility and equity outcomes. The paper proceeds as follows. Section 2 introduces the data and methodological setup, detailing scenario definitions, travel demand generation, epidemic simulation, and the optimization framework. Section 3 presents the results, moving from epidemic dynamics to aggregate, temporal, and spatial accessibility impacts, and concludes with policy implications. Section 4 summarizes key insights and outlines directions for epidemic-resilient transport planning.

\section{Data and Methods}

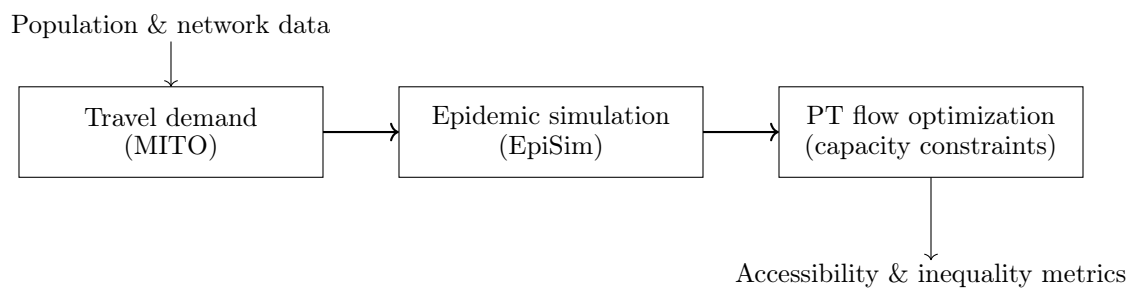
\begin{figure}[ht]
\centering
\begin{tikzpicture}[font=\small]

\draw (0,0) rectangle (4,1.2);
\node at (2,0.8) {Travel demand};
\node at (2,0.4) {(MITO)};

\draw (5,0) rectangle (9,1.2);
\node at (7,0.8) {Epidemic simulation};
\node at (7,0.4) {(EpiSim)};

\draw (10,0) rectangle (14,1.2);
\node at (12,0.8) {PT flow optimization};
\node at (12,0.4) {(capacity constraints)};

\draw[->, thick] (4,0.6) -- (5,0.6);
\draw[->, thick] (9,0.6) -- (10,0.6);

\node at (2,2) {Population \& network data};
\draw[->] (2,1.8) -- (2,1.2);

\node at (12,-1.3) {Accessibility \& inequality metrics};
\draw[->] (12,0) -- (12,-1.1);

\end{tikzpicture}
\caption{Overview of the integrated framework. MITO generates individual travel demand, EpiSim simulates infection dynamics under epidemic measures, and the optimization model evaluates public transport accessibility under capacity constraints.}
\label{fig:framework}
\end{figure}

This section describes the data and methodological framework (see Figure~\ref{fig:framework}) used to analyze the interaction between epidemic dynamics and public transport accessibility. We first introduce the set of policy scenarios under consideration, which define how epidemic measures alter both demand and supply. We then present the data sources and models used for travel demand generation and epidemic simulation. Finally, we outline the optimization-based approach used to evaluate passenger flows under epidemic capacity constraints. This structure ensures that the subsequent analyses can be interpreted with full awareness of both the empirical context and the methodological assumptions. 

\subsection{Scenario Definition}
To evaluate how epidemic dynamics and transport policies interact, we define five epidemic–policy scenarios covering the period from March to May 2020. For consistency across scenarios, the analysis focuses on a representative weekday (Monday), with March 7 serving as the intervention date from which restrictions are applied in all cases. The scenarios differ in the extent to which facilities are closed and in the degree of capacity reduction applied to the public transport (PT) system.

\begin{enumerate}
\item \textbf{Historic scenario.} This scenario replicates the lockdown policies implemented in Germany during March–May 2020 and assumes a 50\% reduction in PT capacity.
\item \textbf{Laissez-faire scenarios.} In these counterfactual settings, no facilities are closed. We distinguish two variants: (a) public transport operates at full nominal capacity (\emph{PT full}), and (b) vehicle capacity is reduced to 50\% (\emph{PT half}).
\item \textbf{10 Percent scenario.} This case represents a mild intervention in which 10\% of all facilities are closed while PT capacity is limited to 50
\item \textbf{Inflection scenario.} This intermediate case assumes that half of all facilities are closed, again combined with a 50\% PT capacity limit.
\item \textbf{Strict scenario.} This setting corresponds to a complete lockdown, with all facilities closed and PT capacity restricted to 50\%.
\end{enumerate}


\begin{table}[t]
\centering
\begin{tabular}{lcccccc}
\hline
\multicolumn{1}{c}{} & Historic & \begin{tabular}[c]{@{}c@{}}Laissez-faire \\ PT Full\end{tabular} &
  \begin{tabular}[c]{@{}c@{}}Laissez-faire \\ PT Half\end{tabular} & 10 Percent & Inflection & Strict \\
\hline
Work           & 20--40\%  & 0\% & 0\%  & 10\% & 50\% & 100\% \\
Education      & 70--100\% & 0\% & 0\%  & 10\% & 50\% & 100\% \\
Nursing        & 70--100\% & 0\% & 0\%  & 10\% & 50\% & 100\% \\
Shopping       & 5--40\%   & 0\% & 0\%  & 10\% & 0\%  & 0\%   \\
Other          & 20--50\%  & 0\% & 0\%  & 10\% & 50\% & 100\% \\
Recreation     & 15--35\%  & 0\% & 0\%  & 10\% & 50\% & 100\% \\
Public Transit & 50\%      & 0\% & 50\% & 10\% & 50\% & 50\%  \\
\hline
\end{tabular}
\caption{Scenario definition by type of facility and percentage restricted during the March--May 2020 period.}
\end{table}

These scenarios are designed to span a broad range of policy settings, from complete absence of restrictions to full lockdown, with intermediate cases reflecting partial facility closures and reduced transport capacity. This design allows us to disentangle the effects of demand suppression and supply constraints, and to assess how different policy combinations shape epidemic and accessibility outcomes.

\subsection{Travel Demand Generation} 
Travel demand for the epidemic simulation was produced using the Microscopic Transportation Orchestrator (MITO), an agent-based travel demand model introduced by~\citet{moeckel2020agentbased}. MITO is an open-source microsimulation tool (\url{https://github.com/msmobility/mito}) that creates individual-level travel demand for each person in the synthetic population of a given study area. In this study, the synthetic population corresponds to the Munich metropolitan region and was generated following the methodology described in~\citet{moreno2018population}. Model calibration was based on data from the German national household travel survey~\citep{infas2018mobilitaet}.  

MITO consists of four modules that generate travel demand: trip generation, destination choice, mode choice, and preferred arrival time. Details on these modules are provided in Appendix~\ref{appendix:mito}. All modules were calibrated against observed data. Furthermore, the generated demand has been validated against data from the Munich Transport and Tariff Association (MVV) in~\citet{tengos_pnas_paper}.  

Because MITO operates at the individual level, various personal characteristics such as vehicle ownership, income, or disability status affect travel choices. To ensure computational efficiency, the synthetic population was reduced to 25\% of its full size for simulation. Data for the local public transport network, including stop locations, routes, trips, and schedules, were obtained from the Germany-wide GTFS feed aggregator \citep{brosi2019gtfs}. Travel times were calculated using the SBB routing module available in MATSim \citep{sbb2020matsimextensions}, which is based on GTFS transit information. Access and egress portions of trips were simulated using MATSim’s walking model on a road network derived from OpenStreetMap \citep{osm}.  

\begin{figure}[t]
	\centering
	\begin{subfigure}[b]{0.23\textwidth}
		\centering		
		\includegraphics[trim={1.5cm 1.5cm 1.5cm 1.5cm},clip,height=3cm]{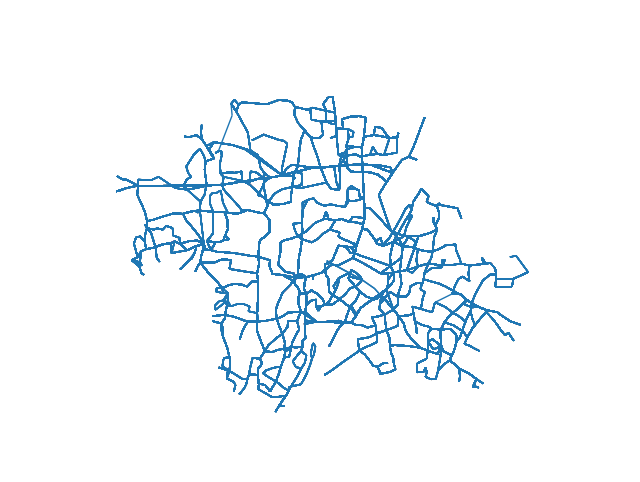}
		\caption{Bus}
	\end{subfigure}
	\hfill
	\begin{subfigure}[b]{0.23\textwidth}
		\centering
		\includegraphics[trim={1.5cm 1.5cm 1.5cm 1.5cm},clip,height=3cm]{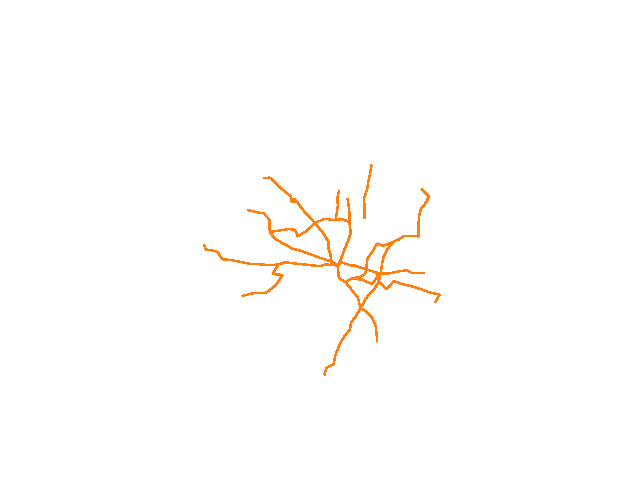}
		\caption{Tram}
	\end{subfigure}
    \hfill
	\begin{subfigure}[b]{0.23\textwidth}
		\centering
		\includegraphics[trim={1.5cm 1.5cm 1.5cm 1.5cm},clip,height=3cm]{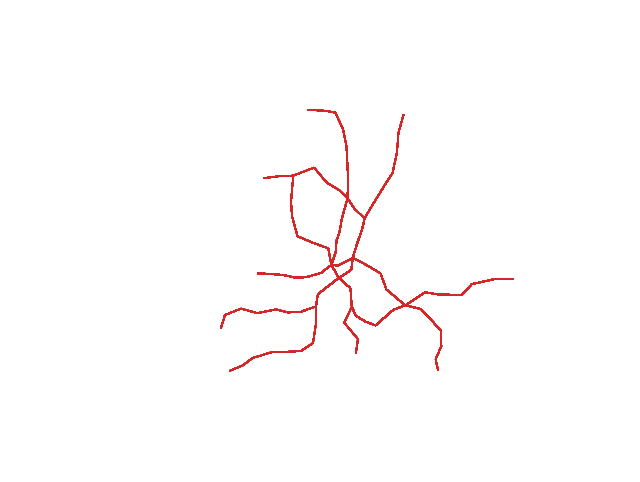}
		\caption{Subway}
	\end{subfigure}
	\hfill
    \begin{subfigure}[b]{0.23\textwidth}
		\centering		
		\includegraphics[trim={1.5cm 1.5cm 1.5cm 1.5cm},clip,height=3cm]{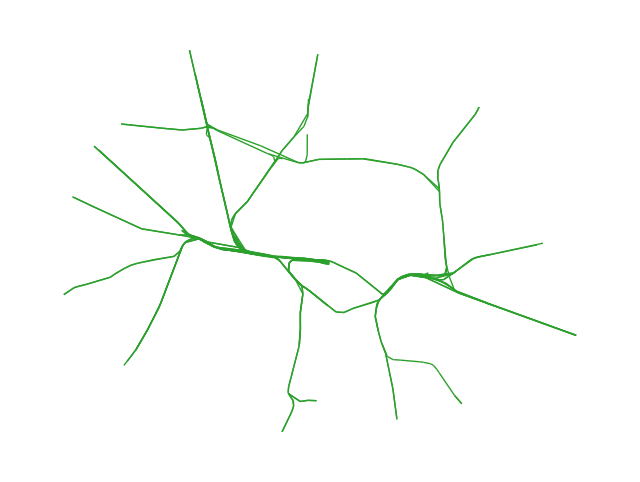}
		\caption{Rail}
	\end{subfigure}
	\caption{Public transport network of Munich used in the study, separated by mode (Bus, Tram, Subway, Rail). The network representation is based on GTFS schedule data and provides the spatial and modal framework for the travel demand generation and subsequent epidemic--transport analyses.}
	\label{fig:experimental-design:munich-network}
\end{figure}

\subsection{Epidemic Simulation}
The epidemic spread simulator EpiSim takes as input the demand from MITO and the MATSim simulation. EpiSim consists of three coupled models: a contact model, an infection model, and a disease progression model \citep{Muller2020}. The contact model specifies who interacts with whom at a given location (in transit or at an activity facility). For each contact, the infection model calculates a probability of transmission based on contact intensity, contact duration, viral shedding, and intake. If a person is infected, the disease progression model determines their transition between states (exposed, infectious, recovered). The simulation runs for one year or until no further infections occur. For details, see~\citet{Muller2020}.  

The Munich metropolitan scenario was calibrated following~\citet{mueller2021}. The key calibration parameter is $\theta$ in the infection probability equation:

\begin{equation} \label{eq:1}
p(\text{infect}\mid \text{contact})_{n,t} = 1 - \exp\left(-\theta \sum_{m} sh_{m,t} \cdot ci_{nm,t} \cdot in_{n,t} \cdot \tau_{nm,t}\right). 
\tag{3.1 \cite{mueller2021}}
\end{equation}

Calibration accounts for activity participation reduction (from Google’s COVID-19 mobility reports), school closures, mask mandates, social distancing, contact tracing, weather effects, and disease imports. The model was fitted to case and hospitalization numbers from the Robert Koch Institute for the first 100 days of the epidemic in the study region (February 16--May 26, 2020).  

\subsection{Combinatorial Optimization Approach (CO)}
\label{section:methodology:co}

\newcommand{\mPathsOnlyMIP}[0]{\ensuremath{\text{MIP}_{\lambda}}}

We formulate the problem of evaluating passenger flow in an epidemic setting as a multi-commodity flow problem with epidemic-aware constraints on a multilayered graph.
Let $\mGraph = (\mNodes, \mArcs)$ be the temporal and spatial expansion of the transportation network consisting of nodes $\mNodes = \mNodesRouteLayer \cup \mNodesWaitingLayer$ and arcs $\mArcs = \mArcsRouteLayer \cup \mArcsWaiting \cup \mArcsTransition \cup \mArcsWalking$. Each transit line schedule is decomposed into transit routes $r \in \mSetTransitRoutes$, where a single route represents a vehicle operating the line, differentiated only by their starting time. Herein, the transit route layer consists of nodes $\mNodesRouteLayer$ representing stops visited at time $t$ and arcs $\mArcsRouteLayer$ connecting them according to their respective schedule. Next, $\mNodesWaitingLayer$ consists of nodes of stops in the public network, passengers' origins and destinations, whereas arcs $\mArcsWaiting$ represents walking between these nodes. Finally, $\mArcsTransition$ and $\mArcsWaiting$ connects these layers, by either connecting nodes of $\mNodesRouteLayer$ and $\mNodesWaitingLayer$ through transition arcs ($\mArcsTransition$), or connect temporal-expanded stops to represent waiting ($\mArcsWaiting$). 

We define $\mArcCapacity_{ij}^r$ as the nominal capacity of transit route $r$, assigned to each corresponding arc $(i,j) \in \mSetArcs$. During an epidemic event, we assume countermeasures are to be implemented to reduce infections by spatial distancing. This effectively reduced the available capacity of transit route vehicles. 

Let $p \in \mSetPassengers$ be a tuple consisting of the origin-destination of passengers and their respective start time and maximum travel time. Each origin and destination is connected to the network by arcs representing walking to and from a station to enter the network. Note that these arcs only connect to nodes reachable in space -- one kilometer away -- and time.

Given a set of available routes, passengers are routed through the network along a feasible path $l \in \mSetPaths_p$. Let $\lambda_{l}^{p}$ be one if path $l \in \mSetPaths_p$ is the selected path of passenger $p \in  \mSetPassengers$ and zero otherwise. Furthermore, let $(y_l^p)_{ij}$ be 1 if the arc $(i,j)\in\mSetArcs$ is part of path $l \in \mSetPaths$. Using this notation, we calculate the total travel time of passenger $p$ using path $l$ with
\begin{align}
    \mSum{l \in L_p}{} \lambda_{l}^{p} \mSum{(i, j) \in \mSetArcs}{} c_{ij} (y_l^p)_{ij}
\end{align}

The objective is to select a set of routes to minimize the maximum cumulative flow over any arc, effectively aiming to minimize the exposure of passengers to each other and reducing epidemic spread. 
Herein, the mathematical formulation is as follows.

\begin{align}
    \min_{\bm{\lambda}} \mSum{p \in \mSetPassengers}{} \mSum{l \in L_p}{} \lambda_{l}^{p} \mSum{(i, j) \in \mSetArcs}{} c_{ij} (y_l^p)_{ij} \label{eq:objective}
\end{align}
s.t.
\begin{align}
\mSum{p \in \mSetPassengers}{} \mSum{l \in \mSetPaths_p}{} (y_l^p)_{ij} \lambda_{l}^{p} &\leq \mArcCapacity_{ij} 
& \forall (i,j) \in \mSetArcs^r, r \in \mSetTransitRoutes \label{eq:constr:capacity}\\
\mSum{l \in \mSetPaths_p}{} y_l^p &= 1 
& \forall p \in \mSetPassengers \label{eq:constr:convexity}\\
\lambda_{l}^{p} &\in \{0,1\} 
& \forall p \in \mSetPassengers, \forall l \in \mSetPaths_p \label{eq:vars:paths}\\
\end{align}

The objective is defined in \eqref{eq:objective}, where we select a path $l \in \mSetPaths$ for each passenger $p$ to minimize the total cost of routing all passengers in $\mSetPassengers$. Constraint~\eqref{eq:constr:capacity} ensures that the capacity of the transit route $r$ is never exceeded. Constraint~\eqref{eq:constr:convexity} states that exactly one path per passenger should be selected.  
Finally, Constraint~\eqref{eq:vars:paths} defines the domain of the decision variables.

\paragraph{Price and branch solver}
We devised our resolution approach based on the work of \cite{LienkampSchiffer2024}, where they showed that the multi-commodity flow problem, originating from assigning passengers to paths, can be solved using a \gls{pnb} scheme. 
It was designed for networks with up to 56,295 passengers for a two hour time frame in the morning (7am-9am), whereas this study we consider an 4 times increase in demand, with around 218,528 passengers to be routed and considering the whole-day demand and network. This substantial increase in problem size required an efficient re-implementation and improvements through means of efficient data structures and parallelization. Details on the pricing approach, as well as speed-up techniques applied, are described in Appendix~\ref{appendix:pnb}.

\subsection{Discussion}  
The analysis focuses on passenger flows within the City of Munich. Nevertheless, part of the demand naturally originates outside the region or is destined beyond its boundaries. To account for this exogenous demand without explicitly modeling the surrounding networks, we estimate the share of such trips from the aggregated daily demand for each mode. We assume that this share corresponds to half of the nominal vehicle capacity being occupied by exogenous flows. Accordingly, the effective capacity available for routing within Munich is reduced in all experiments. This assumption is consistent with common practice in disruption and vulnerability studies, where exogenous demand is incorporated through adjusted capacity parameters rather than by explicitly modeling external networks.  

A further methodological consideration concerns the use of an optimization-based approach for assigning passenger flows, instead of a full agent-based simulation. We emphasize that these approaches are complementary rather than competing. While epidemic dynamics are modeled microscopically using EpiSim, the optimization model provides a tractable and transparent way to quantify system-level accessibility under epidemic capacity constraints. Its role is not to replicate individual route-choice behavior, but to establish a conservative benchmark: if unmet demand arises even under optimal routing, it would likely be at least as severe in decentralized, real-world settings. Moreover, recent evidence shows that at an aggregate level, the flow patterns resulting from optimization-based and simulation-based assignments are highly similar, supporting the validity of this approach~\citep{tengos_pnas_paper}.  

Taken together, these assumptions and methodological choices enable us to capture both the behavioral dynamics of epidemic spread and the structural constraints of the transport system in a computationally feasible way. With this framework in place, we next evaluate the defined scenarios to analyze how epidemic restrictions affect accessibility across aggregate, temporal, and spatial dimensions.

\section{Results}
Our results unfold in four steps, moving from epidemic dynamics to transport outcomes. We first validate the epidemic simulation and examine how restrictions reshape infection environments (Section~\ref{subsec:epidemicValidation}). We then quantify aggregate accessibility losses (Section~\ref{subsec:aggregatedResults}), before disaggregating them over time of day (Section~\ref{subsec:temporalResults}) and across space (Section~\ref{subsec:spatialResults}). This progression from mechanisms to aggregate, temporal, and spatial effects reveals both the scale and distribution of accessibility challenges under epidemic conditions. Section~\ref{subsec:synthesis} synthesizes these findings into overarching insights and policy implications for epidemic-resilient transport planning.

\subsection{Epidemic Validation and Dynamics}\label{subsec:epidemicValidation}
\begin{figure}[!ht]
    \centering
    \includegraphics[width=0.7\textwidth]{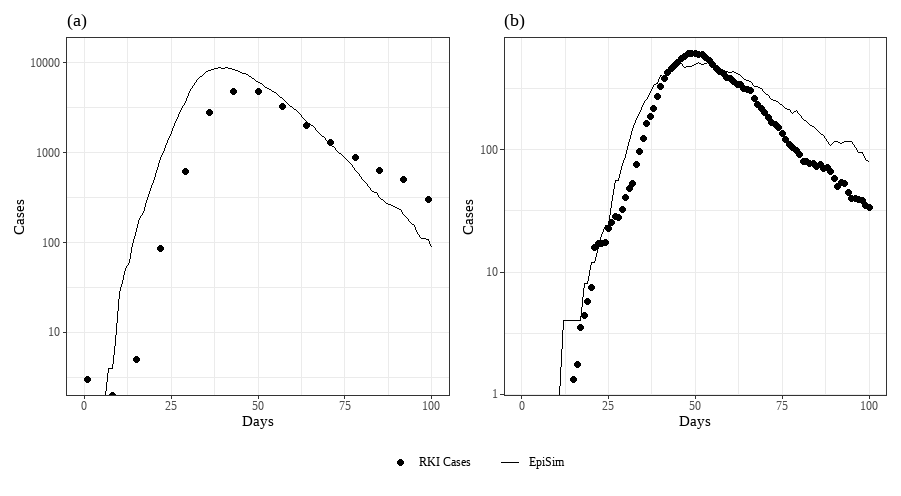}
    \caption{Validation of epidemic dynamics: Comparison of simulated epidemic trajectories (EpiSim) with official statistics from the Robert Koch Institute (RKI). (a) Daily infections reproduce the epidemic curve but peak earlier and at higher levels, consistent with under-reporting in official data. (b) Daily hospitalizations align closely with observed values.}
    \label{fig:episimVsRKI}
\end{figure}

Figure \ref{fig:episimVsRKI} compares epidemic trajectories from EpiSim to official statistics from the \gls{rki}. Panel (a) presents daily infections, while panel (b) shows daily hospitalizations. The simulated curves capture the characteristic progression of the epidemic: rapid growth, a pronounced peak around day 50, and a subsequent decline. This correspondence provides confidence that the epidemic model realistically represents the dynamics relevant for the transport analysis. Some deviations are observable and instructive. The infection peak in the simulation occurs slightly earlier than in the \gls{rki} data and at a higher level. This effect can be explained with under-reporting during the early pandemic phase, when testing capacities were limited and reporting lags were common. In contrast, hospitalizations, which are less prone to under-reporting, align much more closely between the simulation and observed data. 
Taken together, these results suggest that the model may somewhat overestimate early infections but overall provides a robust approximation of epidemic pressures. This validation step is critical, since the credibility of the transport analysis depends on epidemic dynamics being well captured.

\begin{figure}[!ht]
    \centering
    \includegraphics[width=0.7\textwidth]{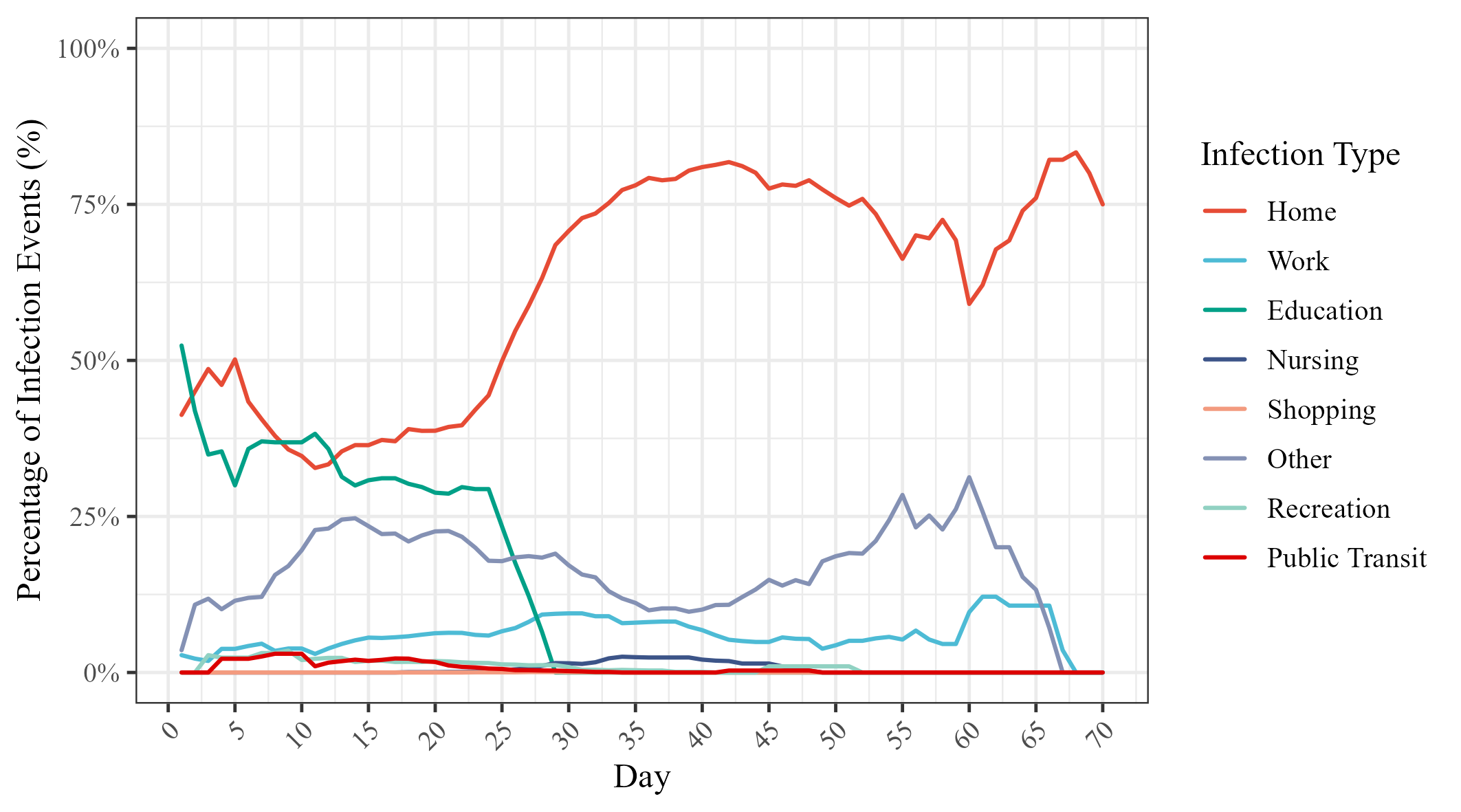}
    \caption{Infection events by day by activity type, 7-day moving average: Daily infections disaggregated by activity setting. Before restrictions, workplaces, schools, and leisure activities contribute substantially to infection events. After restrictions, transmission shifts predominantly to households. The figure shows how epidemic measures redistribute rather than eliminate infection risks.}
    \label{fig:infectionByDayByType}
\end{figure}

Beyond validation, our EpiSim model reveals how infections are distributed across activity types. Figure \ref{fig:infectionByDayByType} shows the daily number of infection events differentiated by households, workplaces, schools, and leisure activities. Before restrictions, infections occur across all settings, with workplaces and schools playing a prominent role. Once restrictions are introduced, external infection channels drop sharply, and transmission becomes overwhelmingly concentrated within households. This shift highlights a central point: epidemic interventions not only reduce the overall number of cases but also change where infections occur. Public transport itself is not identified as a dominant infection driver in the simulation. Instead, its relevance arises indirectly through its role in enabling or constraining access to activities. This implies that the importance of transport measures lies not primarily in preventing in-vehicle transmission but in shaping demand and accessibility patterns during epidemics.

\subsection{Macroscopic Analysis}\label{subsec:aggregatedResults}
\begin{figure}[!ht]
	\centering
	\includegraphics[width=0.6\textwidth]{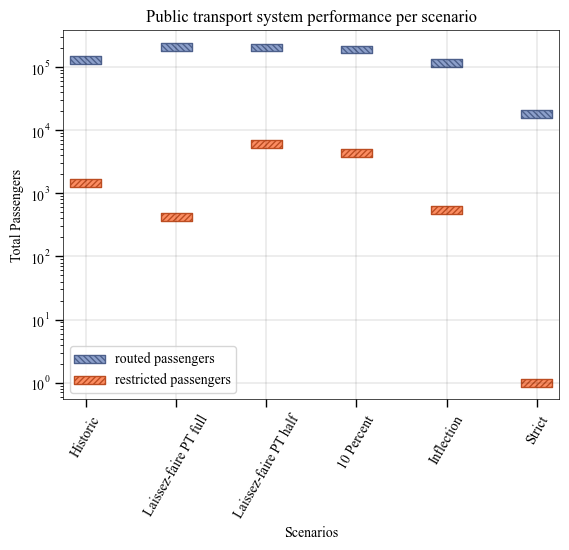}
	\caption{Total number of routed and restricted passengers across scenarios. Laissez-faire settings with halved PT capacity generate severe mismatches, while facility closures reduce demand and alleviate bottlenecks. The figure illustrates the nonlinear interaction of epidemic measures and transport restrictions in shaping accessibility.}
	\label{fig:aggregated:passenger-demand}
\end{figure}

Figure~\ref{fig:aggregated:passenger-demand} presents, for each scenario, the total number of passengers that can be routed through the public transport network as well as those who remain restricted because their trips cannot be accommodated within available capacity. This macroscopic view highlights the fundamental trade-off between epidemic mitigation and mobility provision.
The laissez-faire scenarios provide an illustrative starting point. When no facilities are closed and demand remains high, the system is congested once capacity restrictions are applied. In the PT full case, most demand can still be routed, but once vehicle capacity is halved (PT half), the number of restricted passengers rises dramatically. In this setting, tens of thousands of passengers are effectively excluded from public transport on a single weekday. The mismatch between supply and demand illustrates the vulnerability of PT systems under uncoordinated epidemic interventions: if demand is left unchecked while supply is cut, accessibility crises emerge.

In contrast, scenarios that combine facility closures with capacity limits produce very different outcomes. The 10\% closure scenario reduces total demand just enough to relieve the congestion effect in the system. While some passengers are still restricted, the gap between routed and restricted demand is substantially narrower than in the laissez-faire PT half scenario. This pattern becomes even clearer in the inflection scenario, where 50\% of facilities are closed. Here, demand falls sharply, and the number of restricted passengers declines to relatively modest levels. The strict scenario demonstrates the extreme: with nearly all facilities closed, overall travel demand collapses, and very few passengers are restricted — but only because very few passengers are traveling at all.

The historic scenario lies between these extremes. It reflects the actual combination of partial facility closures and a 50\% PT capacity limit that was implemented in Munich during the first wave of COVID-19. The results show that this policy mix managed to suppress demand enough to avoid a catastrophic mismatch, but still left a non-negligible number of passengers restricted.

Two important insights emerge from this aggregate analysis. First, epidemic and transport policies interact in nonlinear ways. Restricting transport capacity without suppressing demand (as in laissez-faire PT half) produces severe bottlenecks, while moderate facility closures can substantially mitigate accessibility losses. Second, epidemic resilience cannot be achieved by looking at supply-side or demand-side measures in isolation. Instead, the overall outcome depends on their combination. This highlights the need for coordinated policy design: modest demand suppression can sometimes substitute for large supply cuts, achieving similar epidemic protection at a far lower social cost in terms of mobility.

\subsection{Temporal accessibility and bottlenecks}\label{subsec:temporalResults}
\begin{figure}[!ht]
	\centering
	\begin{subfigure}[b]{0.32\textwidth}
		\centering
		\includegraphics[height=3.8cm]{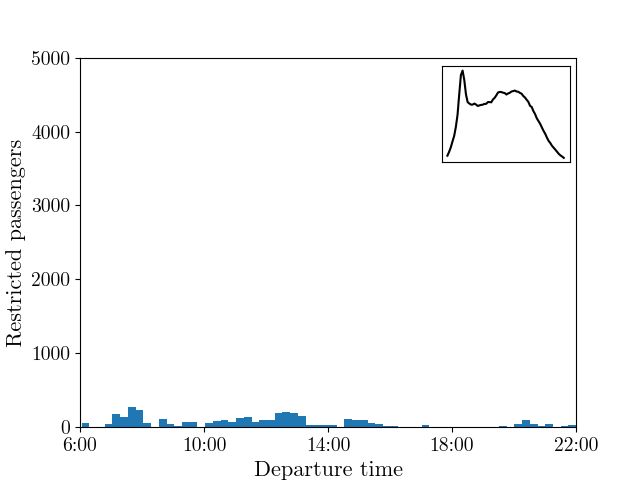}
		\caption{Laissez-faire PT full}
    	\label{fig:results:temporal-analysis:lf-full}
	\end{subfigure}
    \begin{subfigure}[b]{0.32\textwidth}
		\centering
		\includegraphics[height=3.8cm]{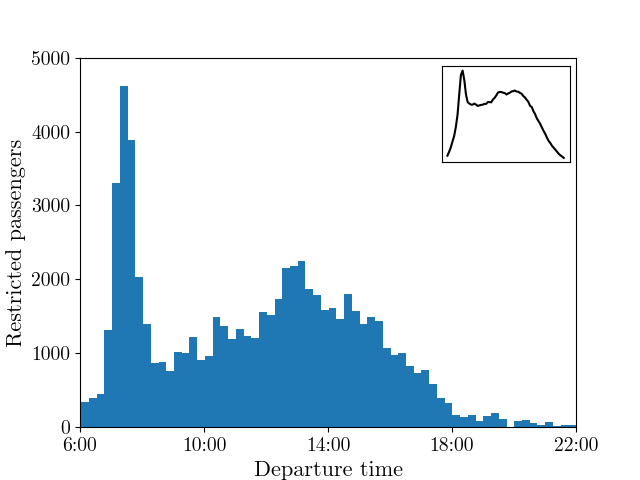}
		\caption{Laissez-faire PT half}
        \label{fig:results:temporal-analysis:lf-half}
	\end{subfigure}
	\begin{subfigure}[b]{0.32\textwidth}
		\centering
		\includegraphics[height=3.8cm]{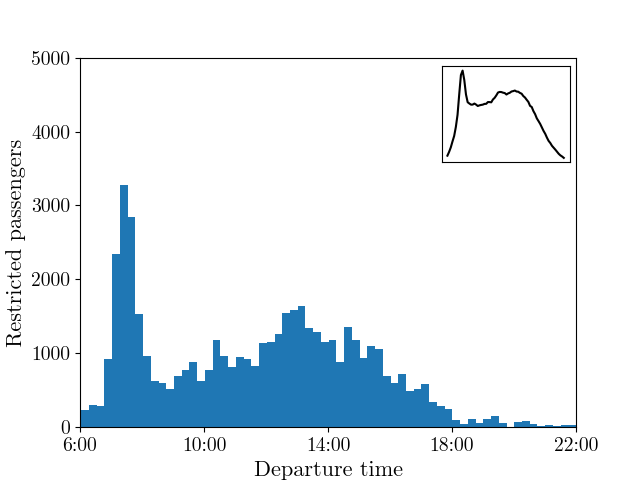}
		\caption{10\% facilities closed}
        \label{fig:results:temporal-analysis:10pct}
	\end{subfigure}

    \begin{subfigure}[b]{0.48\textwidth}
		\centering
		\includegraphics[height=3.8cm]{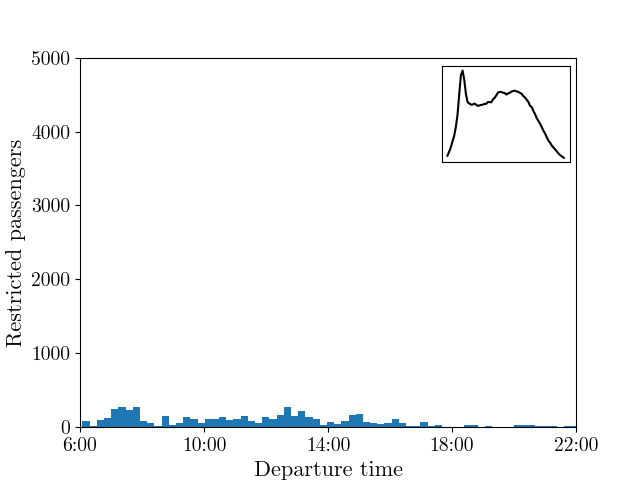}
        \caption{Inflection Scenario}
        \label{fig:results:temporal-analysis:infl}
	\end{subfigure}
    \begin{subfigure}[b]{0.48\textwidth}
		\centering
		\includegraphics[height=3.8cm]{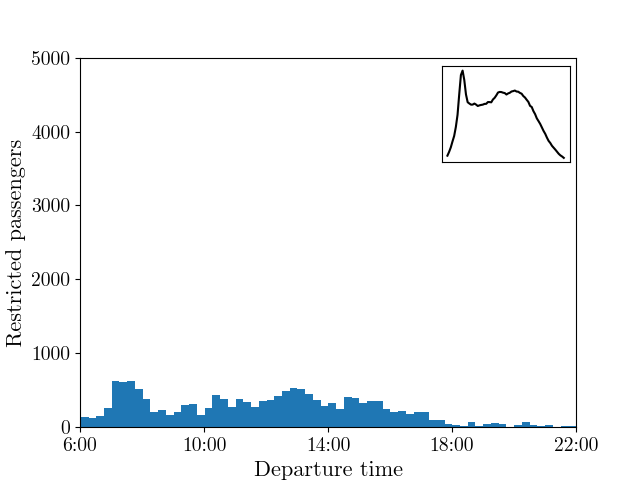}
		\caption{Historic Scenario}
        \label{fig:results:temporal-analysis:historic}
	\end{subfigure}
    \caption{Restricted passengers by time of day across scenarios, with the baseline demand profile shown as an inset. Accessibility problems are concentrated in morning and afternoon peaks, while evenings and nights remain largely unaffected. The figure highlights how epidemic pressures amplify pre-existing peak-hour vulnerabilities in public transport systems.}
	\label{fig:results:temporal-analysis}
\end{figure}

Aggregate figures conceal an important dimension: when during the day accessibility losses occur. Figure~\ref{fig:results:temporal-analysis} therefore disaggregates restricted passengers by time of day for each scenario, with the baseline demand curve shown as an inset for comparison. This analysis highlights how epidemic and transport policies interact with the natural rhythms of urban mobility.

Across all scenarios, the results show that accessibility problems are concentrated during the morning (6:30–9:00) and afternoon (13:30–18:00) peaks. These periods correspond to commuting hours, when demand is highest and capacity is most strained. Epidemic restrictions thus do not create entirely new vulnerabilities but instead amplify existing ones: peak hours, already critical under normal conditions, become the primary points of failure in the system. The midday period (9:00–13:30) also shows significant unmet demand in high-demand scenarios, particularly in laissez-faire settings. 
By contrast, the evening (18:00–22:00) and night (22:00–6:30) hours remain largely unaffected. Even under tight capacity constraints, the number of restricted passengers during these times is negligible. This asymmetry across the day underscores that uniform restrictions are inefficient. Limiting capacity throughout the day imposes social costs without addressing the actual pressure points, which are highly concentrated in time.

Comparing across scenarios reveals further insights. In the laissez-faire PT full setting, restrictions remain moderate, but when vehicle capacity is halved (laissez-faire PT half), the mismatch during peak hours becomes extreme, with more than 60,000 restricted passengers at midday. The 10\% and inflection scenarios, which incorporate facility closures, suppress demand sufficiently to reduce the scale of bottlenecks, though they do not eliminate them. The historic scenario once again sits between these extremes, producing fewer restrictions than laissez-faire PT half but still leaving a significant share of peak-hour travelers unable to access the system. Finally, the strict scenario nearly eliminates restricted passengers, but only by virtually eliminating demand — a solution that is clearly unsustainable from a mobility perspective.

Taken together, these temporal results highlight two key points. First, epidemic pressures exacerbate the structural vulnerabilities of PT systems: peak-hour congestion becomes the dominant source of accessibility loss. Second, time-differentiated policies hold great potential. Instead of blanket restrictions, measures such as staggered work start times, school scheduling adjustments, or targeted fleet reinforcements during peaks could substantially reduce unmet demand while keeping restrictions light in less critical periods.


\subsection{Spatial Inequalities}\label{subsec:spatialResults}
\begin{figure}[!htb]
	\centering
	\includegraphics[width=1\textwidth]{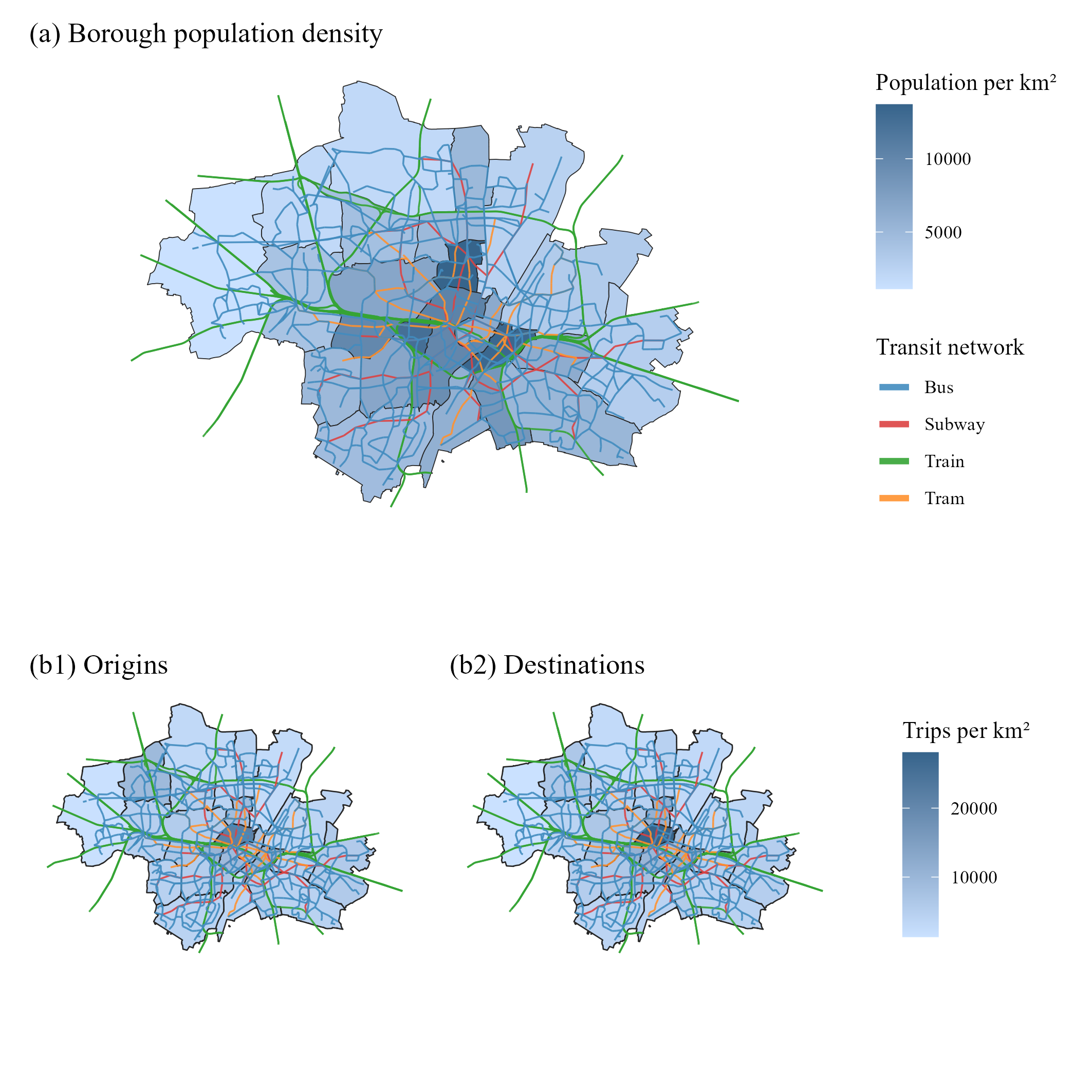}
	\caption{Population density by borough: Spatial context for the transport analysis. (a) Population density is highest in central Munich. (b) Trip demand density shows dispersed origins but highly concentrated central destinations. The figure underscores the dependence of peripheral boroughs on reliable public transport connections.}
	\label{fig:pop_density}
\end{figure}

\begin{figure}[!htb]
	\centering
	\includegraphics[width=1\textwidth]{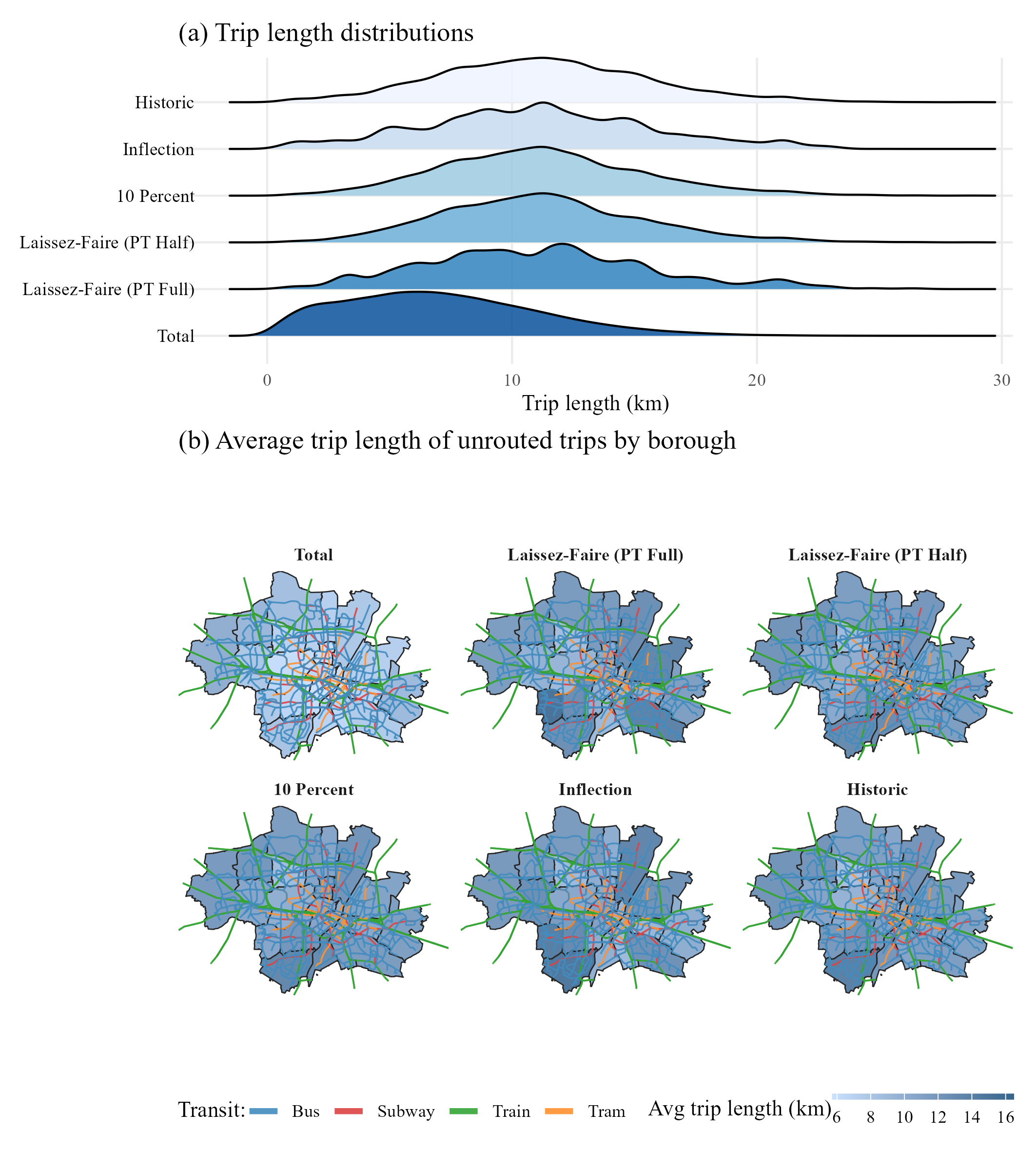}
	\caption{Comparison of trip lengths for unrouted trips across scenarios, with 'Total' denoting the average trip length including all trips. (a) Distribution of unrouted trips is skewed toward longer distances relative to the baseline. (b) Average origin trip lengths of unrouted passengers are highest in peripheral boroughs. The figure demonstrates that long-distance commuters face disproportionate accessibility losses under epidemic restrictions.}
	\label{fig:trip_length_dist_trip_length_unrouted_borough}
\end{figure}

\begin{figure}[!htb]
	\centering
	\includegraphics[width=0.9\textwidth]{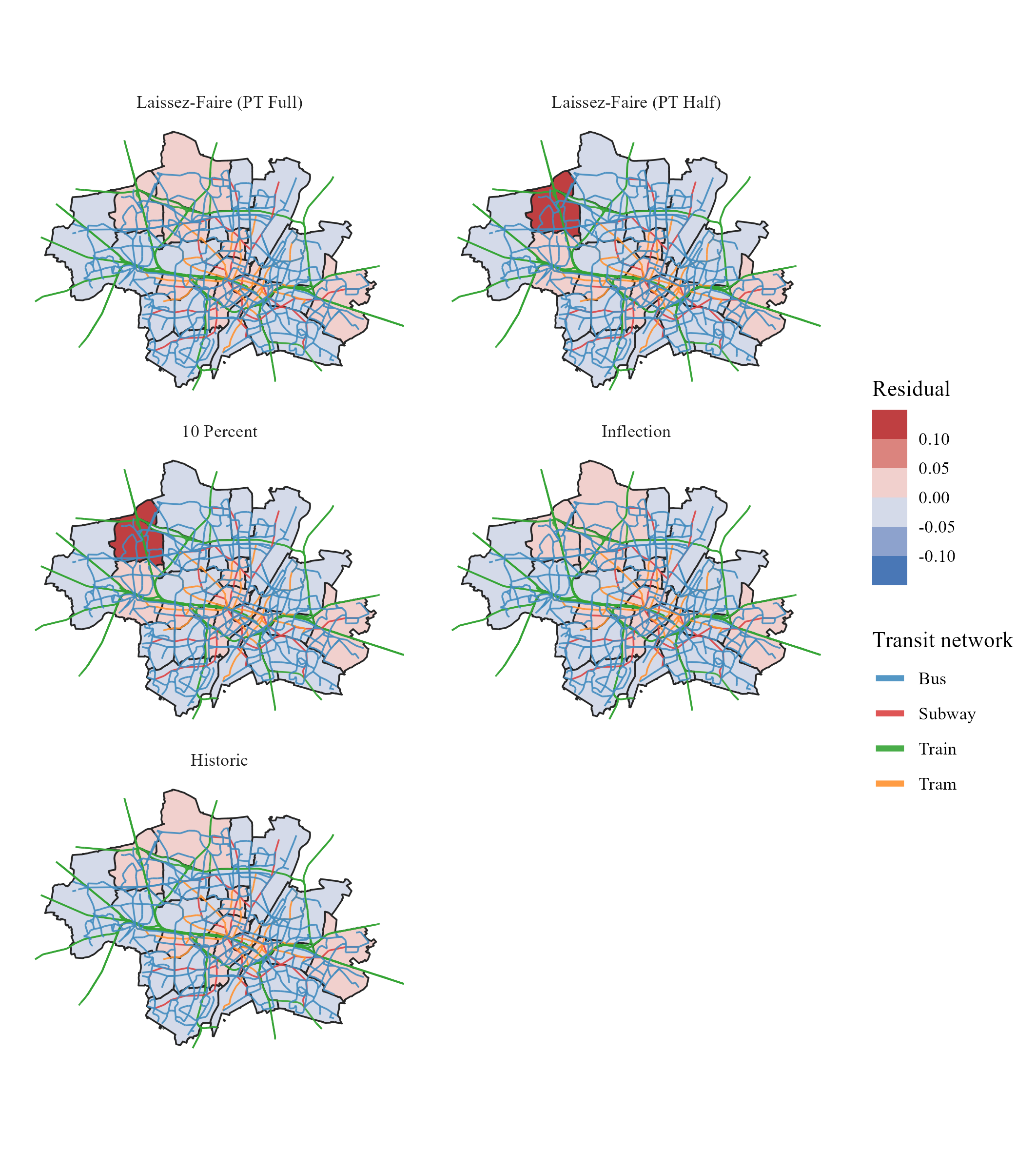}
\caption{ Residuals of unrouted trip shares after controlling for borough population density. The residuals are calculated with $r$ being the unrouted trip share and $D$ population density. Fit $r = \alpha + \beta \log(1+D) + \varepsilon$. Red areas indicate excess burdens beyond what density would predict, while blue areas perform better than expected. The figure reveals that structural disadvantages, particularly in peripheral boroughs, drive inequalities in accessibility.}

	\label{fig:unrouted_trips_density_residual}
\end{figure}

\begin{figure}[]
	\centering
	\begin{subfigure}[b]{0.48\textwidth}
		\centering
		\includegraphics[trim={1.5cm 1.5cm 1.5cm 1.5cm}, clip, height=6cm]{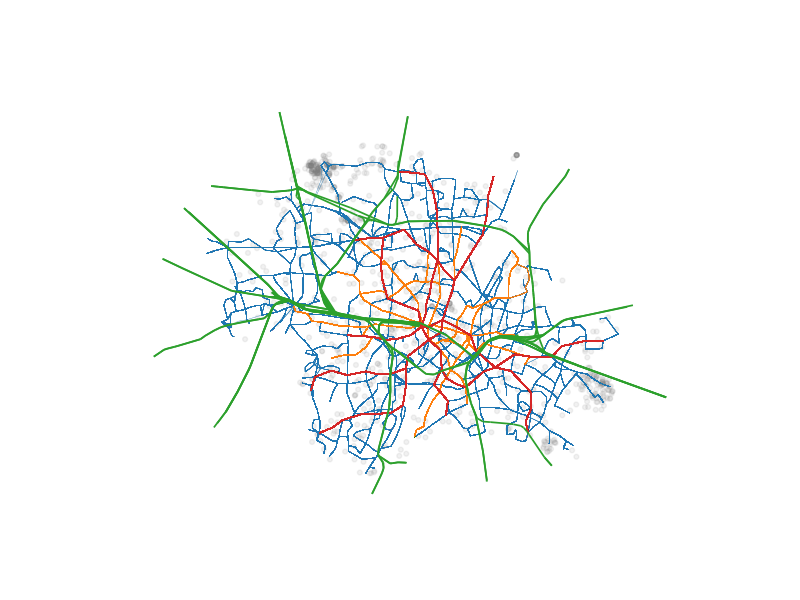}
        \caption{Laissez-faire PT full}
	\end{subfigure}
    \hfill
    \begin{subfigure}[b]{0.48\textwidth}
		\centering
		\includegraphics[trim={1.5cm 1.5cm 1.5cm 1.5cm}, clip, height=6cm]{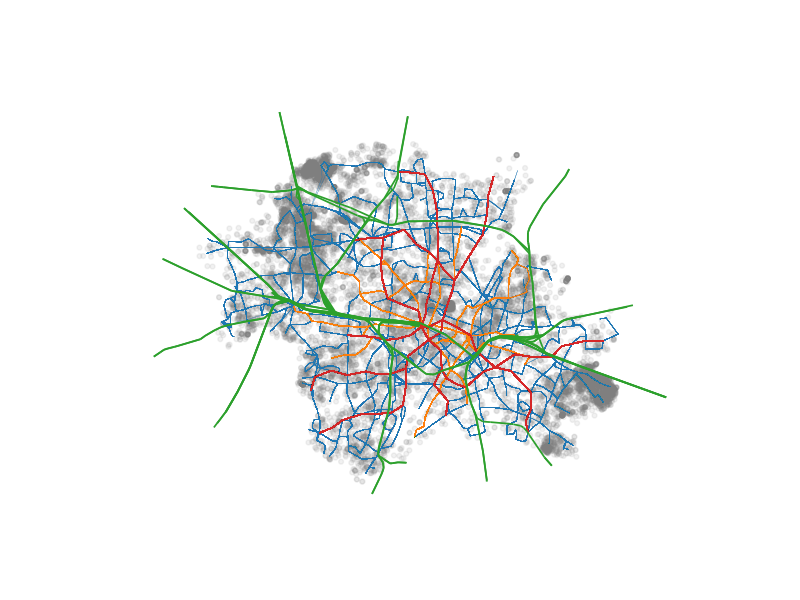}
        \caption{Laissez-faire PT half}
	\end{subfigure}
	\begin{subfigure}[b]{0.48\textwidth}
		\centering
		\includegraphics[trim={1.5cm 1.5cm 1.5cm 1.5cm}, clip, height=6cm]{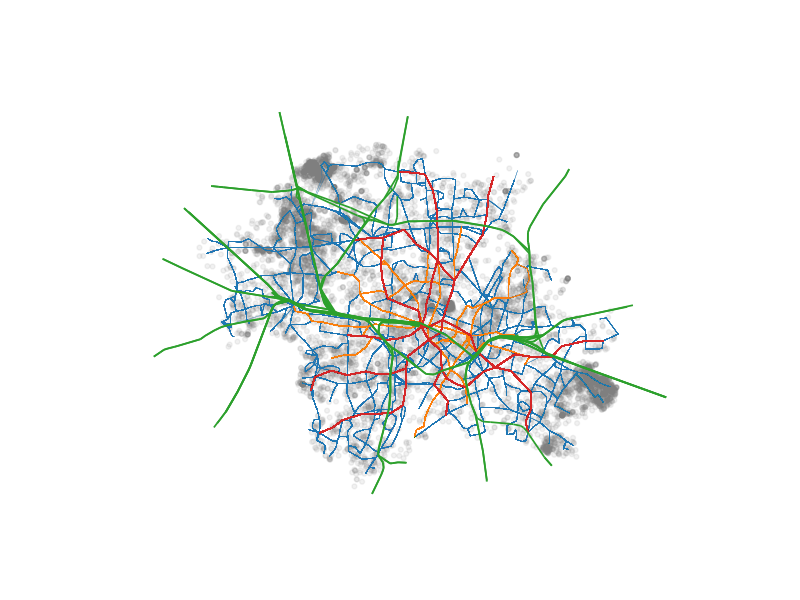}
		\caption{10\% facilities closed}
	\end{subfigure}
    \hfill
    \begin{subfigure}[b]{0.48\textwidth}
		\centering
		\includegraphics[trim={1.5cm 1.5cm 1.5cm 1.5cm}, clip, height=6cm]{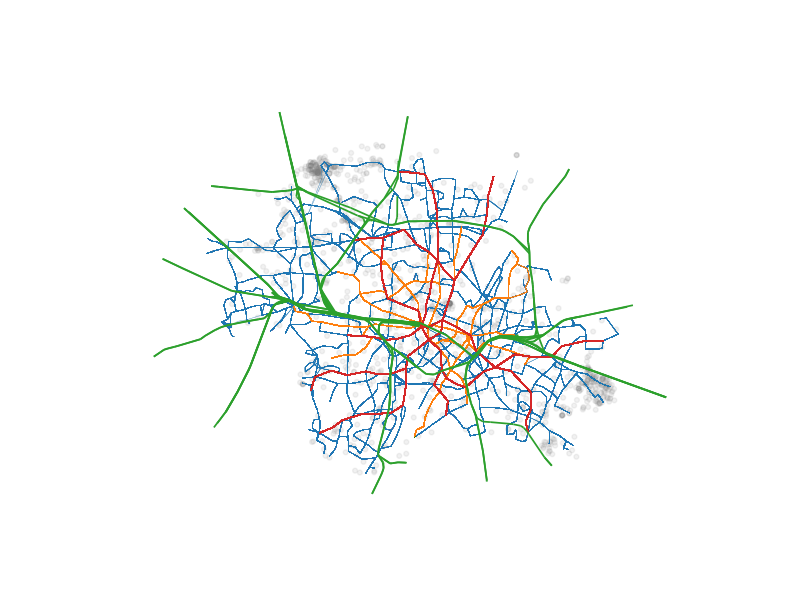}
		\caption{Inflection Scenario}
	\end{subfigure}
    \begin{subfigure}[b]{0.48\textwidth}
		\centering
		\includegraphics[trim={1.5cm 1.5cm 1.5cm 1.5cm}, clip, height=6cm]{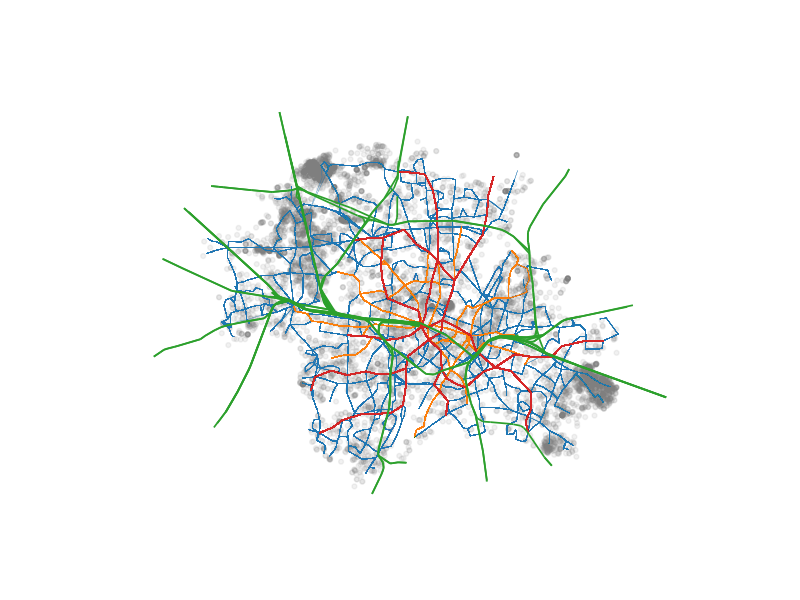}
		\caption{Historic Scenario}
	\end{subfigure}
    \hfill
    \begin{subfigure}[b]{0.48\textwidth}
		\centering
		\includegraphics[height=6cm]{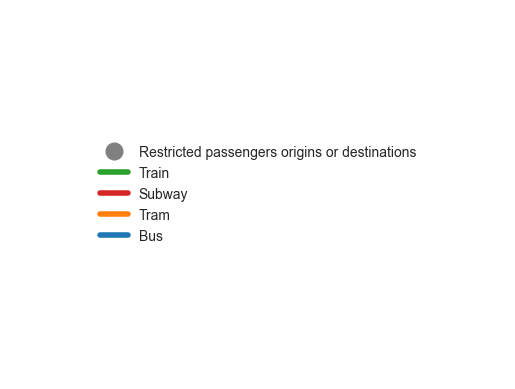}
	\end{subfigure}
	\caption{Restricted passengers by borough across scenarios. (a) Laissez-faire PT full, (b) Laissez-faire PT half, (c) 10 Percent, (d) Inflection, (e) Historic. Accessibility losses are concentrated in peripheral areas, especially under laissez-faire PT half. The figure illustrates how different epidemic-policy combinations shape the geography of accessibility.}
	\label{fig:results:spatial-analysis}
\end{figure}

The temporal analysis highlights when accessibility problems are most severe. Equally important, however, is where these problems occur. Epidemics rarely affect all parts of a city equally, and transport restrictions can exacerbate pre-existing inequalities between central and peripheral areas. Figures~\ref{fig:pop_density}--\ref{fig:results:spatial-analysis} present the spatial perspective, linking population density, demand patterns, and scenario-specific accessibility outcomes.

Figure \ref{fig:pop_density} provides the baseline context. Panel (a) shows population density, which is highest in the city center and declines toward the periphery. Panel (b) depicts trip demand density, revealing a different pattern: while many trips originate in outer boroughs, their destinations cluster strongly in central areas. This asymmetry underscores the structural dependence of peripheral residents on reliable public transport for commuting to central employment, education, and service hubs.

The implications of this dependence become visible in Figure \ref{fig:trip_length_dist_trip_length_unrouted_borough}, which compares the length of unrouted trips to the baseline distribution and shows borough-level averages. Restricted trips are disproportionately longer than average, and peripheral boroughs consistently exhibit the highest average origin trip lengths for unrouted passengers. In other words, those who already travel farther are the most likely to be excluded under epidemic restrictions. This finding highlights a double burden for peripheral residents: longer commutes and higher risk of unmet demand.

To test whether these inequalities are simply a function of population density, Figure \ref{fig:unrouted_trips_density_residual} presents density-adjusted residuals. Even after controlling for density, several peripheral boroughs show substantial positive residuals, meaning that their accessibility losses are higher than expected. This indicates that inequities are not merely demographic but reflect structural disadvantages in the transport network — for example, fewer alternative modes, lower service frequencies, or greater reliance on single corridors.

Figure~\ref{fig:results:spatial-analysis} then compares spatial accessibility outcomes across scenarios. In the laissez-faire PT half scenario, accessibility losses are most severe, and the outer boroughs bear the brunt of restrictions. Facility closure scenarios, particularly the inflection case, reduce both overall demand and the severity of inequalities, though at the cost of suppressing mobility across the board. The historic scenario again lies between extremes: outer areas still suffer disproportionate losses, but less than under laissez-faire half. The strict scenario effectively eliminates inequities, but only by collapsing mobility system-wide.

Two broader insights emerge. First, epidemic restrictions deepen existing spatial divides. Central boroughs, with dense populations and better service coverage, retain comparatively higher accessibility, while peripheral areas face exclusion. Second, the severity of inequalities is not fixed but shaped by policy choices. Coordinated demand-supply strategies can reduce inequities, whereas uniform capacity cuts without demand suppression (as in laissez-faire half) make them worse.


\subsection{Implications}\label{subsec:synthesis}
The preceding analyses have shown how epidemics affect transport systems along several dimensions: they alter epidemic dynamics across activities, generate aggregate accessibility losses, produce temporal bottlenecks, and amplify spatial inequalities. While each of these findings is informative in its own right, their true value emerges when considered together. This section therefore synthesizes the results into three overarching insights that cut across all scenarios: first, how epidemic restrictions reshape the distribution of risks; second, how epidemic and transport policies interact in nonlinear and sometimes counterintuitive ways; and third, how these dynamics reinforce temporal and spatial inequalities in accessibility. We then translate these insights into policy implications for designing epidemic-resilient and equitable transport systems.

\paragraph{Synthesis}
The combined evidence from the validation, activity-based epidemic dynamics, aggregate transport impacts, temporal accessibility, and spatial inequalities provides a comprehensive picture of how epidemics and transport systems interact. Rather than treating these dimensions in isolation, it is important to integrate them to understand systemic trade-offs. In the following, we distill the results into three overarching insights that cut across all scenarios: (i) how epidemic restrictions shift the distribution of risks, (ii) how transport and epidemic policies interact in nonlinear ways, and (iii) how these dynamics amplify existing temporal and spatial inequalities. This structure guides the discussion and prepares the ground for identifying concrete policy implications.

First, epidemic restrictions change not only the scale but also the distribution of risks.
A central lesson from the epidemic dynamics is that interventions do not simply flatten the curve; they reallocate where infections occur. Our simulations demonstrate that household transmission becomes dominant once external contacts are curtailed. This means that the rationale for restricting transport lies less in preventing direct in-vehicle infections and more in shaping broader mobility patterns. Transport measures should therefore be understood as structural levers: they influence who can access essential activities, when, and at what cost. This is a subtle but important reframing, as it moves the debate from “does public transport spread infection?” to “how do transport policies mediate epidemic and social outcomes simultaneously?”.

Second, epidemic and transport policies interact in nonlinear ways.
Our aggregate and temporal analyses reveal that the impact of transport restrictions depends strongly on the surrounding policy environment. Cutting vehicle capacity without reducing demand leads to severe mismatches, while even modest demand suppression through facility closures can alleviate pressure on the network. In other words, transport and epidemic policies are not additive but interdependent: their combination determines whether accessibility bottlenecks emerge or are contained. This implies that the effectiveness of interventions cannot be judged in isolation — coordinated policies matter. A modest reduction in demand can sometimes substitute for large supply cuts, achieving similar health protection at much lower social cost.

Third, epidemic pressures amplify existing temporal and spatial inequalities.
Epidemics do not affect all travelers equally, and our results show that restrictions reinforce long-standing fault lines in urban accessibility. Morning and afternoon peaks, already critical stress points, become the periods with the greatest exclusion. Peripheral boroughs, where residents have longer commutes and fewer modal alternatives, suffer disproportionate accessibility losses. These inequities are not explained solely by population density; they reflect deeper structural disadvantages in the transport network. Similarly, workers and larger households — groups with less flexibility to reduce trips — are more heavily affected. Epidemic restrictions thus risk reinforcing socio-spatial divides in access to essential services.

Taken together, these findings suggest that uniform restrictions are both inefficient and inequitable. Blanket capacity limits applied throughout the day and across the entire network impose unnecessary costs while failing to address the times and places where accessibility problems are most severe.

\paragraph{Policy implications}
Our integrated analysis highlights several directions for more effective and equitable epidemic transport policies. Crucially, the results show that transport interventions cannot be designed in isolation: they must be coordinated with broader epidemic strategies, and they must explicitly address when and where accessibility losses occur.

\begin{description}
\item[Time-differentiated measures.] Peak-hour bottlenecks emerge as the primary source of unmet demand. Uniform restrictions across the day create unnecessary costs because evenings and nights show little strain on the system. Policies such as staggered work start times, flexible school hours, or temporary fleet reinforcements targeted to the morning and afternoon peaks can reduce exclusion while minimizing disruption at other times. This implies that epidemic transport management is less about reducing total capacity and more about smoothing demand over time.
\item[Spatially informed capacity management.] Inequalities are not evenly distributed across the network. Peripheral boroughs and long-distance commuters face disproportionately high burdens, often with limited access to alternatives such as walking or cycling. Supporting these areas through additional services, reserved capacity for essential workers, or subsidies for alternative modes can prevent epidemic measures from exacerbating existing socio-spatial divides. Here, transport authorities can build on equity principles already familiar from accessibility planning, but apply them dynamically under epidemic constraints.
\item[Coordinated demand- and supply-side interventions.] The scenarios demonstrate that demand suppression policies (e.g., facility closures, remote-work mandates) can relieve pressure on the transport system more effectively than supply cuts alone. Transport authorities should therefore seek coordination with public health measures: capacity reductions in vehicles should go hand in hand with policies that reduce peak travel demand. Without this alignment, accessibility crises are inevitable, as shown in the laissez-faire half-capacity scenario. Conversely, modest demand reductions can substantially mitigate the negative side effects of epidemic restrictions, creating room for less intrusive transport policies.
\item[Equity-focused resilience planning.] Epidemics highlight structural weaknesses in urban mobility systems, particularly for vulnerable groups such as low-income households without cars, shift workers with inflexible schedules, and residents of the periphery. Future preparedness strategies should explicitly identify these groups and incorporate equity metrics into resilience planning. This could mean prioritizing their access in capacity allocation, or building redundancy into networks to protect them against systemic shocks. Equity should thus not be treated as a secondary concern but as a core criterion for evaluating epidemic interventions.
\end{description}

Taken together, these implications point to a broader lesson: epidemic resilience in transport requires a move away from blanket restrictions toward targeted, flexible, and equity-aware policies. Time- and space-sensitive measures, combined with strong coordination across policy domains, offer a path to simultaneously protect public health, maintain mobility, and safeguard fairness.

\section{Conclusion}

This study examined how epidemic dynamics and public transport systems interact, focusing on the trade-offs between health protection, mobility provision, and social equity. By integrating agent-based epidemic simulation with an optimization-based passenger flow model, we developed a framework that captures both behavioral epidemic dynamics and the structural constraints of transport networks. Applying this framework to Munich during the first wave of COVID-19 allowed us to quantify the effects of different policy scenarios on accessibility.  

Three key findings emerge. First, epidemic restrictions alter not only the magnitude but also the distribution of risks: once external activities are curtailed, infections concentrate within households, making public transport policies relevant primarily through their indirect impacts on mobility and accessibility. Second, epidemic and transport policies interact in nonlinear ways: halving vehicle capacity without reducing demand creates severe bottlenecks, whereas even modest demand suppression substantially alleviates accessibility losses. Third, epidemic pressures amplify pre-existing inequalities: accessibility problems are concentrated during commuting peaks and disproportionately affect peripheral boroughs, long-distance commuters, and groups with less flexibility in their travel behavior.  

Taken together, these results highlight that blanket restrictions are both inefficient and inequitable. More effective strategies are targeted, coordinated, and equity-aware. Time-differentiated measures such as staggered work and school schedules, spatially informed interventions to support peripheral areas, and coordinated demand- and supply-side policies offer more balanced outcomes. The framework developed here provides a transparent basis for evaluating such interventions, and can inform epidemic preparedness planning in urban transport systems.  

Future work should extend the analysis in several directions. First, incorporating adaptive behavioral responses in mode and route choice would refine the representation of individual decision-making under epidemic conditions. Second, extending the framework to multimodal networks, including active and shared mobility, would broaden its applicability. Third, evaluating policies under a wider range of epidemic scenarios, including longer time horizons and varying compliance levels, would enhance the robustness of insights. By pursuing these extensions, future research can strengthen the evidence base for designing transport systems that are resilient, equitable, and sustainable in the face of epidemic shocks.

\bibliographystyle{model5-names}
\bibliography{sample}

\FloatBarrier
\newpage
\onehalfspacing
\begin{appendices}
	\normalsize
	
\section{Additional Material}
\label{appendix:material}

\setcounter{figure}{0}
\setcounter{table}{0}
\renewcommand{\thefigure}{\Alph{section}\arabic{figure}}
\renewcommand{\thetable}{\Alph{section}\arabic{table}}

\begin{table}[h]
\begin{tabular}{lcccccc}
\hline
\multicolumn{1}{c}{} & \multicolumn{6}{c}{Scenario}                  \\
\cline{2-7}
Facilities & Historic & \begin{tabular}[c]{@{}c@{}}Laissez-faire \\  PT Full-capacity\end{tabular} &
  \begin{tabular}[c]{@{}c@{}}Laissez-faire \\ PT Half-capacity\end{tabular} &10 Percent & Inflection & Strict \\
\hline
Work           & 20 -- 40\%  & 0\% & 0\%  & 10\% & 50\% & 100\% \\
Education      & 70 -- 100\% & 0\% & 0\%  & 10\% & 50\% & 100\% \\
Nursing        & 70 -- 100\% & 0\% & 0\%  & 10\% & 50\% & 100\% \\
Shopping       & 5 -- 40\%   & 0\% & 0\%  & 10\% & 0\%  & 0\%   \\
Other          & 20 -- 50\%  & 0\% & 0\%  & 10\% & 50\% & 100\% \\
Recreation     & 15 -- 35\%  & 0\% & 0\%  & 10\% & 50\% & 100\% \\
Public Transit & 50\%       & 0\% & 50\% & 10\% & 50\% & 50\%  \\
\hline
\end{tabular}
\caption{Scenario definition by type of facility and percentage restricted during the March - May 2020 period}
\end{table}

\begin{figure}[tbh]
	\centering
	\includegraphics[width=1\textwidth]{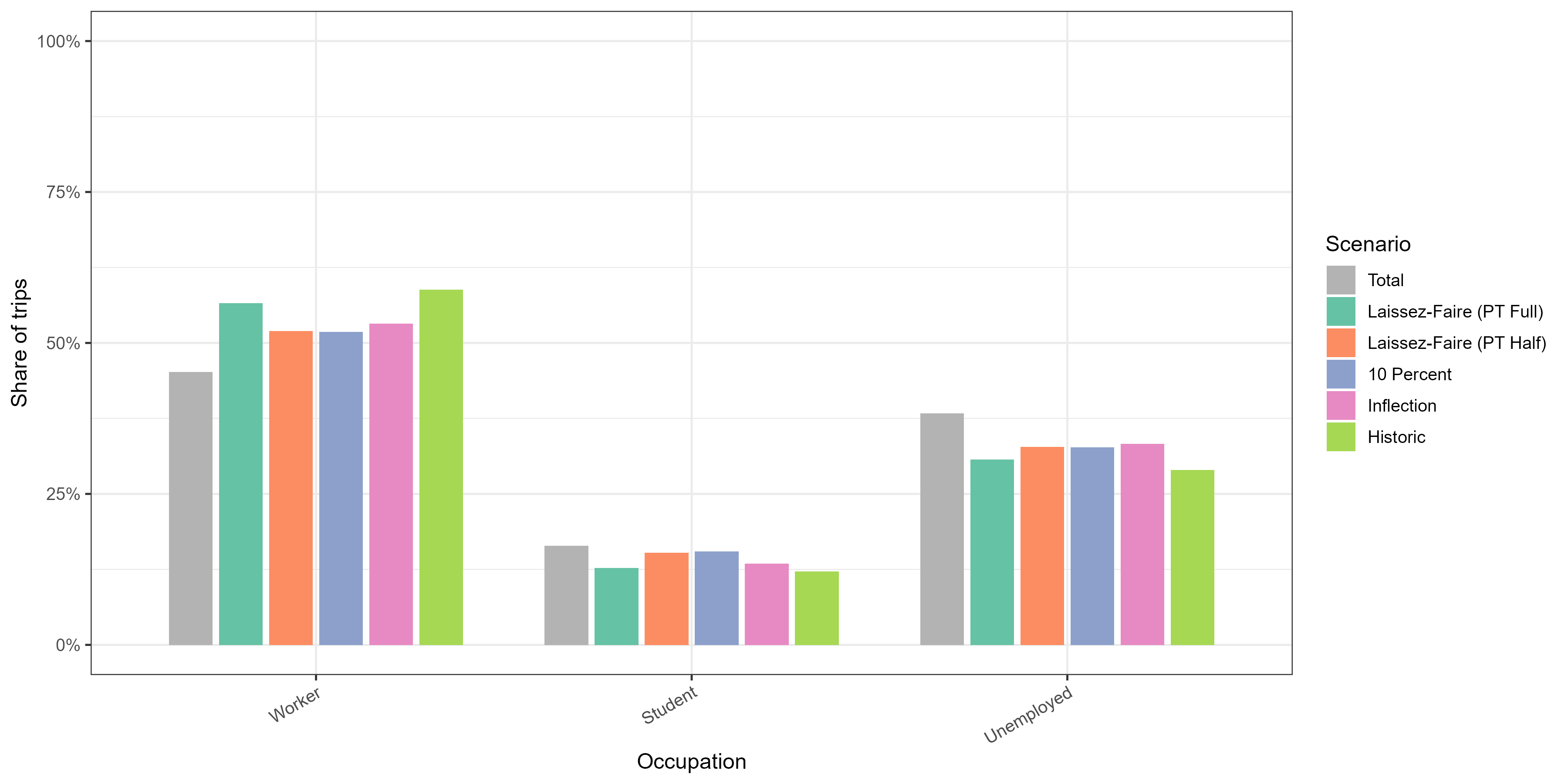}
	\caption{Share of unrouted trips by employment status in each scenario, as compared to the total share of trips by employment status}
	\label{fig:employment_status_bar}
\end{figure}

\begin{figure}[tbh]
	\centering
	\includegraphics[width=1\textwidth]{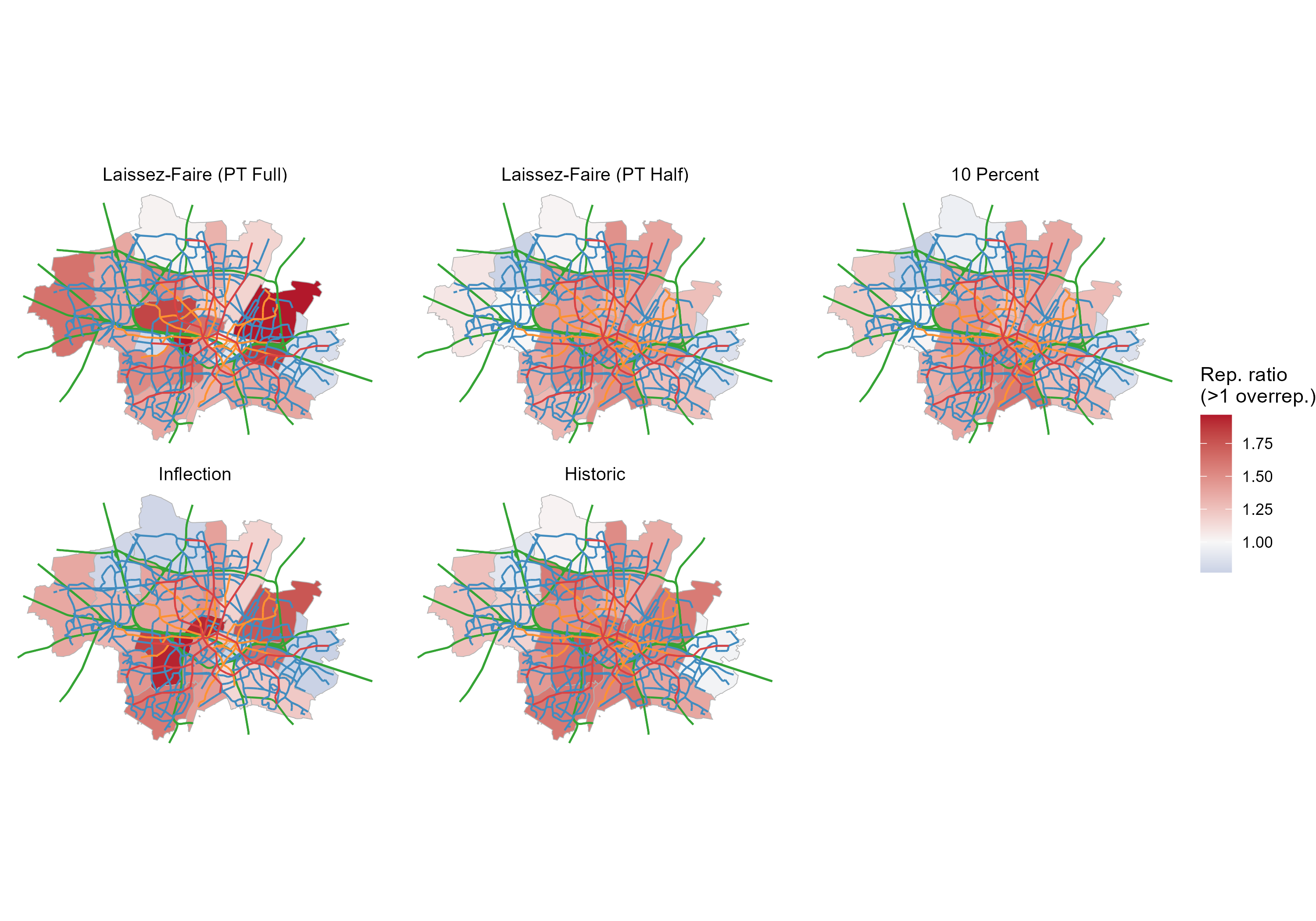}
	\caption{Share of unrouted trips of workers by borough in each scenario}
	\label{fig:employment_status_map}
\end{figure}

\begin{figure}[tbh]
	\centering
	\includegraphics[width=1\textwidth]{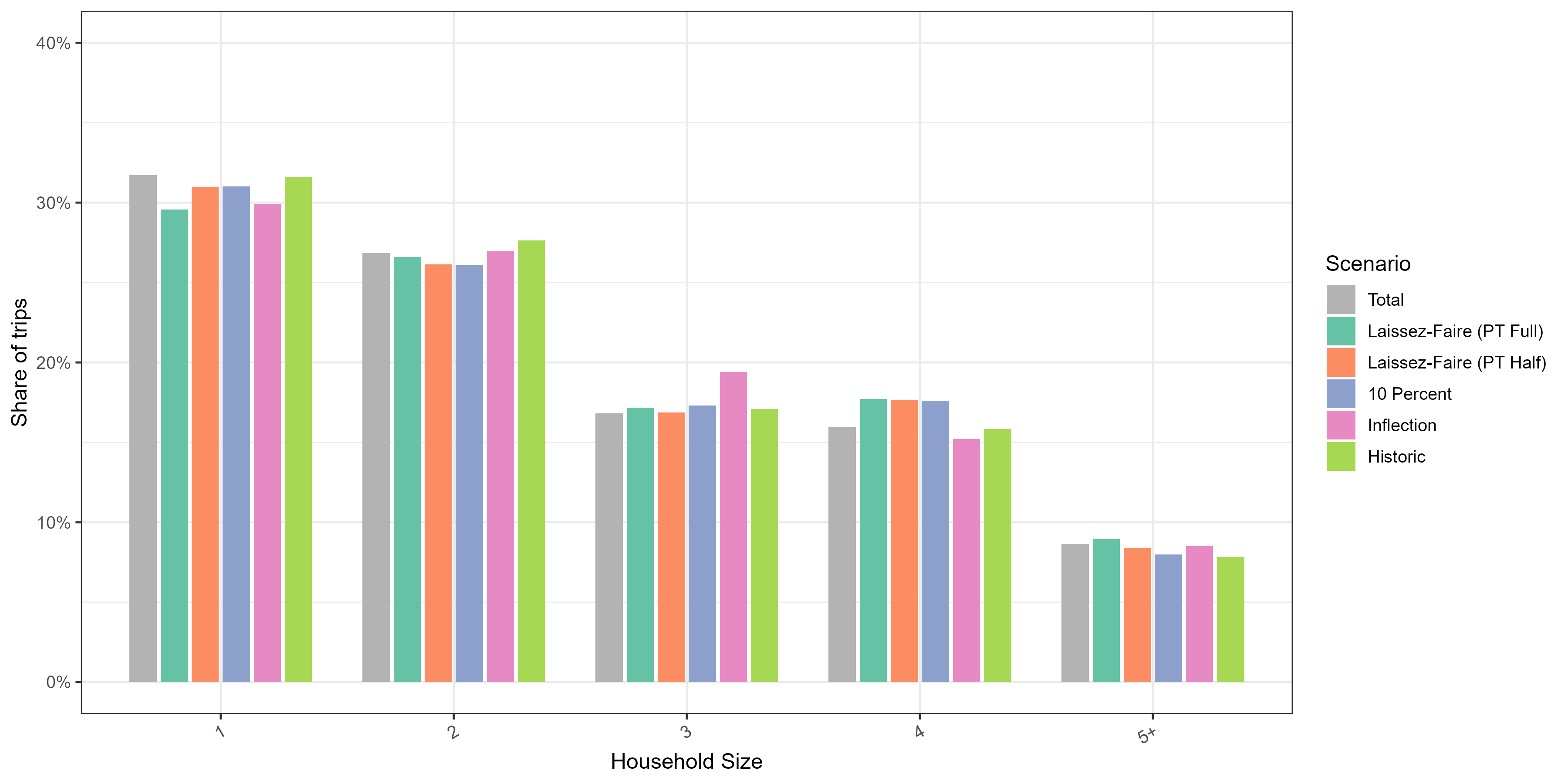}
	\caption{Share of unrouted trips by household size in each scenario, as compared to the total share of trips by household size}
	\label{fig:household_size_bar}
\end{figure}

\begin{figure}[h]
	\centering
	\includegraphics[width=0.5\textwidth]{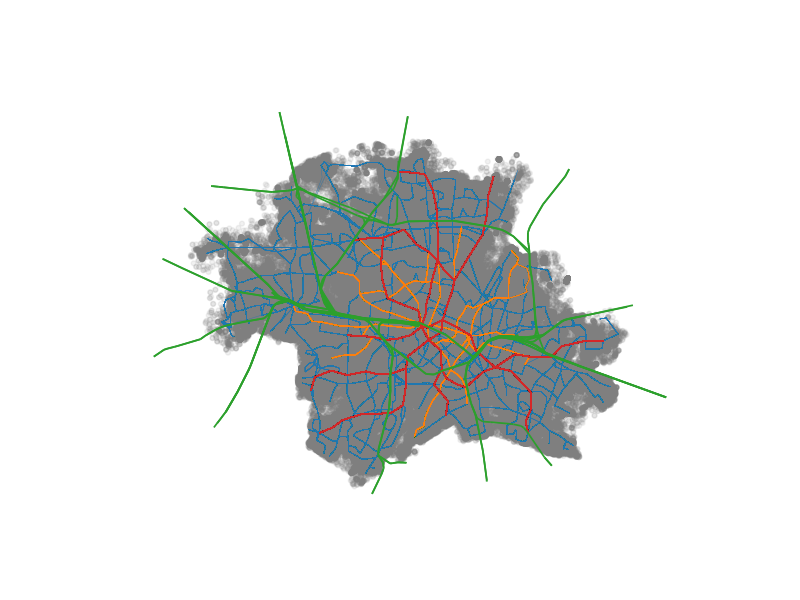}
	\caption{Complete trip demand locations in Laissez-faire PT-Full. The figure illustrates the extend the demand is located which covers most of munich with only small areas in the peripheral with sparser demand.}
	\label{fig:results:spatial-analysis:full_demand}
\end{figure}

\begin{figure}[!htb]
    \centering
    \includegraphics[width=\textwidth]{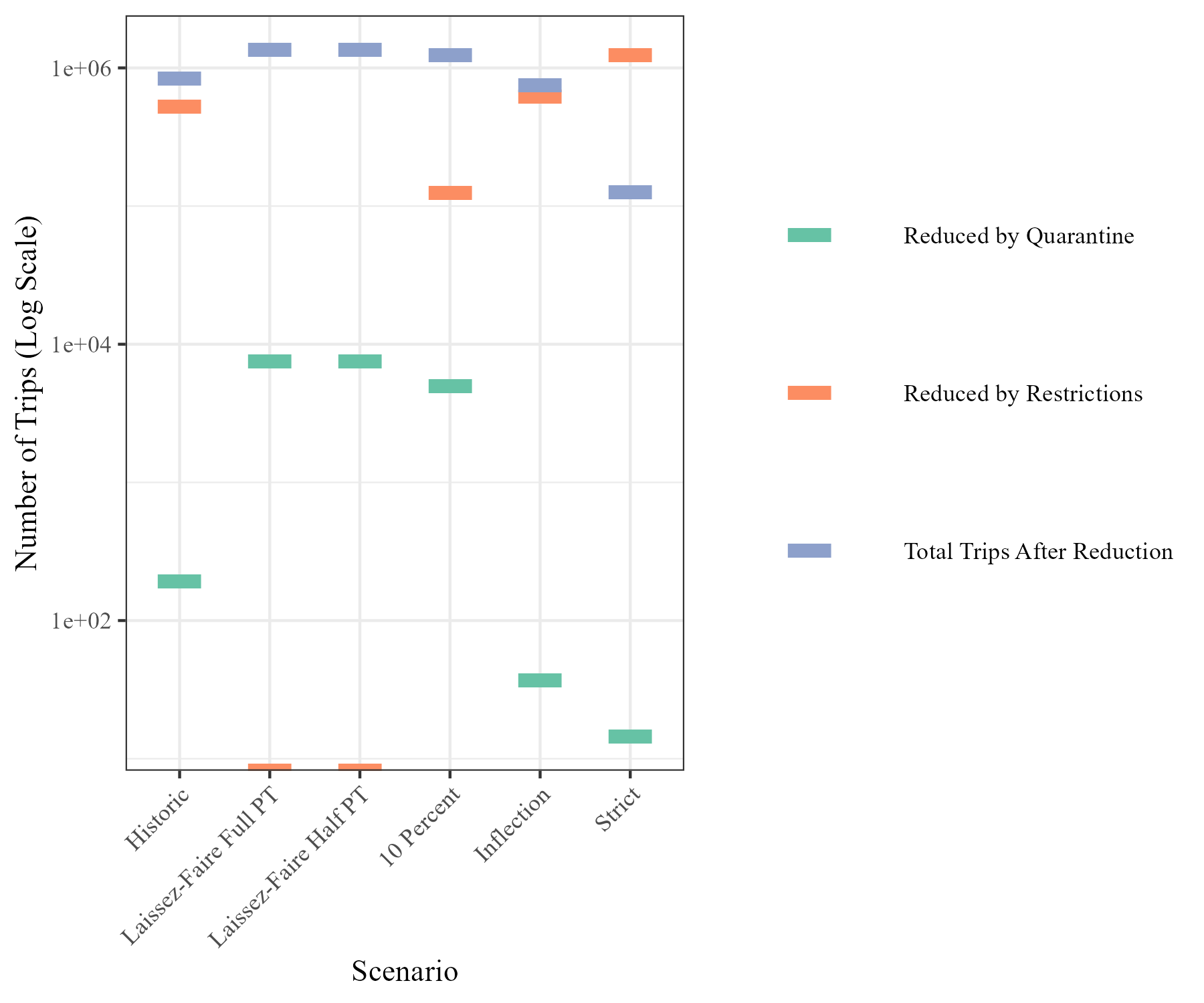}
    \caption{Public transport trip reduction by scenarios}
    \label{fig:pt_trips_reduced_laissezfaire}
\end{figure}


\FloatBarrier

\section{Price-and-branch approach}
\label{appendix:pnb}

In this section, we will summarize additional information on the price-and-branch approach originally proposed in \cite{LienkampSchiffer2024}, additional speed-up techniques implemented for this study, and details on the execution environment.

\paragraph{Pricing filter.} After solving the restricted master problem for a given set of paths, not all paths may need to be recalculated. Based on this idea, we heuristically identify a subset of paths and only solve the pricing problem for these. As a heuristic, we inspect the dual of the capacity constraints. For arcs with negative duals, we identify which passenger flows travel over such arcs and add them to the subset of pricing problems to consider. Note that we still need to evaluate all passengers once at the end. Although the capacity constraints may not be violated, changing flows may have opened the opportunity for passengers to be redirected to a shorter path.

\paragraph{Pricing problem.} In our path-based formulation of the model described in Section~\ref{section:methodology:co}, we proposed a column-generation approach to manage the set of paths per passenger to stay tractable. As shown in previous work, we can formulate a shortest path problem with reduced costs from the informed by the dual solution. This problem can then be solved efficiently using an A* shortest path algorithm, where we use an admissible heuristic based on the shortest path of the non-time-expanded and, therefore, a smaller graph. 

\paragraph{Graph data structure.} The graph in the pricing problem is static, which can be exploited by using an efficient, immutable graph structure. Our A* algorithm searches the graph by iterating all reachable neighbors of any node, i.e., considering only outgoing arcs. Therefore, we encode the graph as continuous list of arcs, sorted by the source of the arc, and a list of ranges per node, indicating which continuous segment of arcs contains its outgoing arcs.

\paragraph{Parallelization.}
In the pricing problem, we iteratively solve an shortest path problem for each passenger to identify a cost reducing path to add to the restricted master problem. As the capacity constraints are not considered at this stage, and the underlying graph is static, we can decompose the problem and solve each shortest path problem independently from each other, leading to a straight-forward parallelization scheme, which directly scales with the number of cpu cores available.

\paragraph{Execution environment.}
We implemented the \gls{pnb} presented in Section~\ref{section:methodology:pb} in the Rust programming language, compiled using rustc version 1.79 in release mode and with link-time optimization. The experiments were conducted on a cluster setup consisting of Intel i9-9900 processors with 16 cores running at 3.10 GHz and 64GB RAM, using Ubuntu 20.04 LTS operating system. We limit our runs to four cores and 32GB of RAM to allow conducting our experiments in parallel. We use Gurobi 9.5 to solve the restricted master problem in the Column Generation algorithm.

\section{Extension: Additional Fleet Assignment}
\label{appendix:extension}

\setcounter{figure}{0}
\setcounter{table}{0}

We extended the model presented in Section~\ref{section:methodology:co} to consider a limited response by the public transit system to capacity bottlenecks by bolstering transit routes, i.e., adding additional vehicles to a transit line, starting at a certain time. 

Herein, let $\mAllocation^r$ be the number of vehicles assigned to a specific transit route $r \in \mSetTransitRoutes$. Let $\mBudget^m$ be the available fleet per transport mode, e.g., bus or subway.

The objective is still to select a set of routes to minimize the maximum cumulative flow over any arc, effectively aiming to minimize the exposure of passengers to each other and reducing epidemic spread. However, the effective capacity of arcs $\mArcCapacity_{ij}$ can be increased by bolstering the corresponding transit route $r \in \mSetTransitRoutes$.
Herein, the mathematical formulation is as follows.

\begin{align}
    \min_{\bm{\lambda}} \mSum{p \in \mSetPassengers}{} \mSum{l \in L_p}{} \lambda_{l}^{p} \mSum{(i, j) \in \mSetArcs}{} c_{ij} (y_l^p)_{ij} \label{eq:ext:objective}
\end{align}
s.t.
\begin{align}
\mSum{p \in \mSetPassengers}{} \mSum{l \in \mSetPaths_p}{} (y_l^p)_{ij} \lambda_{l}^{p} &\leq \mArcCapacity_{ij} \mAllocation^r
& \forall (i,j) \in \mSetArcs^r, r \in \mSetTransitRoutes \label{eq:ext:constr:capacity}\\
\mSum{l \in \mSetPaths_p}{} y_l^p &= 1 
& \forall p \in \mSetPassengers \label{eq:ext:constr:convexity}\\
\mSum{r \in R^m}{} \mAllocation^r &\leq \mBudget^m & \forall m \in \mSetModes \label{eq:ext:constr:budget}\\
\lambda_{l}^{p} &\in \{0,1\} 
& \forall p \in \mSetPassengers, \forall l \in \mSetPaths_p \label{eq:ext:vars:paths}\\
\mAllocation^r &\in \mathbb{N}
& \forall r \in \mSetTransitRoutes \label{eq:ext:vars:allocations}
\end{align}

The objective is defined in \eqref{eq:ext:objective}, where we select a path $l \in \mSetPaths$ for each passenger $p$ to minimize the total cost of routing all passengers in $\mSetPassengers$. Constraint~\eqref{eq:ext:constr:capacity} ensures that the capacity of the transit route $r$ is never exceeded. Constraint~\eqref{eq:ext:constr:convexity} states that exactly one path per passenger should be selected. Constraint~\eqref{eq:ext:constr:budget} limits the total vehicle allocation per mode $m$ within an available budget $\mBudget^m$. Finally, Constraints~\eqref{eq:ext:vars:paths} and \eqref{eq:ext:vars:allocations} define the domains of the decision variables.

\begin{figure}[tb]
	\includegraphics[height=5cm]{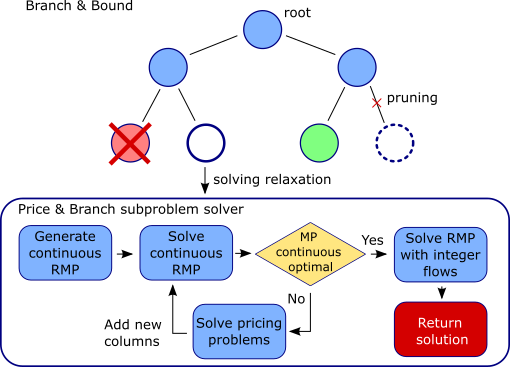}
	\caption{Schematic overview of the branch-and-price-and-branch approach.}
	\label{fig:bpb}
\end{figure}

The path-based model presented in \eqref{eq:ext:objective}--\eqref{eq:ext:vars:allocations} contains two sets of decision variables: i) the path selection $\lambda_{l}^{p}$ for each passenger $p$, and ii) the sizing decision $z^r$ for each transit route $r$. Given a fixed integer assignment of $z^r$, we can use the \gls{pnb} approach to generate all necessary paths and find an integer solution for $\lambda_{l}^{p}$ \citep[c.f.,][]{LienkampSchiffer2024}. We will further refer to this problem as $\mPathsOnlyMIP$. Enumerating all possible sizing configurations is intractable. Therein, we designed an overarching branch-and-bound framework that systematically searches for integer decisions of $z^r$, while we use the \gls{pnb} to find and select paths $\lambda_{l}^{p}$. A schematic overview of this approach can be seen in Figure~\ref{fig:bpb}.

Calculating tight upper bounds helps to accelerate the search by enabling early pruning opportunities. However, preliminary experiments showed that an upper bound is found very late in the search without any additional mechanisms. To this end, we devise a quick heuristic to find potentially sub-optimal but feasible solutions quickly to strengthen the bounding procedure. Algorithm~\ref{alg:heuristic-bounding} shows the corresponding steps. 

We start by inspecting the fractional decisions for $\mAllocation^r$ after solving $\mPathsOnlyMIP$. Let $\left\lfloor \mAllocation^r \right\rfloor$ be the minimum assignment identified. We then define $\overline{\mBudget}^m = \mBudget^m - \mSum{r \in R^m}{} \left\lfloor \mAllocation^r \right\rfloor$ as the residual budget for each mode $m \in \mSetModes$. Iteratively, we now round the allocations with the largest fractional value up, until all residual budget has been used; all remaining allocations are rounded down.
Finally, we solve the $\mPathsOnlyMIP$ with the heuristic allocation decisions $\hat{\mAllocation}$ to find an upper bound solution for the complete problem.

\begin{algorithm}
\caption{Upper bound heuristic}
\label{alg:heuristic-bounding}
\SetKwData{FractionalAllocation}{$\mAllocation$}
\SetKwData{HeuristicAllocation}{$\hat{\mAllocation}$}
\SetKwData{AvailableBudget}{$\mBudget^m$}
\SetKwData{ResidualBudget}{$\overline{\mBudget}^m$}
\SetKwFunction{argmax}{${\bf argmax}$}
\SetKwFunction{solve}{solve\_\mPathsOnlyMIP\_with\_allocation}
\KwIn{Fractional allocation decisions $\FractionalAllocation^r$; Available Budgets $\AvailableBudget$}
\For{$m \in \mSetModes$}{
	$\ResidualBudget \leftarrow \AvailableBudget - \mSum{r \in R^m}{} \left\lfloor \FractionalAllocation^r \right\rfloor$\;
	$\hat{\mRoutes}^m \leftarrow \mRoutes^m$\;
	\While{$\ResidualBudget > 0$}{
		$r' \leftarrow \argmax_{r \in \hat{\mRoutes}^m}\{ \FractionalAllocation^r - \left\lfloor \FractionalAllocation^r \right\rfloor \}$\;
		$\HeuristicAllocation^{r'} \leftarrow \left\lceil \FractionalAllocation^{r'} \right\rceil$\;
		$\hat{\mRoutes}^m \leftarrow \hat{\mRoutes}^m \setminus \{r'\}$\;
		$\ResidualBudget  \leftarrow \ResidualBudget  - 1$\;
	}
	\For{$r' \in \hat{\mRoutes}^m$}{$\HeuristicAllocation^{r'} \leftarrow \left\lfloor \FractionalAllocation^{r'} \right\rfloor$\;}
}
\Return \solve($\HeuristicAllocation^r$)
\end{algorithm}

Note that in case of $\mSum{r \in R^m}{} \left\lceil \mAllocation^r \right\rceil \leq \mBudget^m, \forall m \in \mSetModes$, we can find a trivial integer solution by rounding all fractional allocations decisions up and use the passenger path decisions from the $\mPathsOnlyMIP$ solution without any need to reevaluate.

After applying the upper bound heuristic, we still need to continue with our branch-and-bound procedure by deciding on the branching scheme. Herein, we first identify transit route modes where the budget is tight, i.e., allocations can not be trivially be rounded up. For these modes, we select the most fractional allocation $\mAllocation$, i.e., where $\left\lvert \mAllocation^r - \left\lfloor \mAllocation^r \right\rfloor - 0.5\right\rvert$ is minimal.
We search the tree node by node, always considering the node with the smallest lower bound, as defined by its parent, first. 

\subsection{Allocation and reliability analysis.}
\begin{figure}[tb]
	\centering
	\includegraphics[height=6cm]{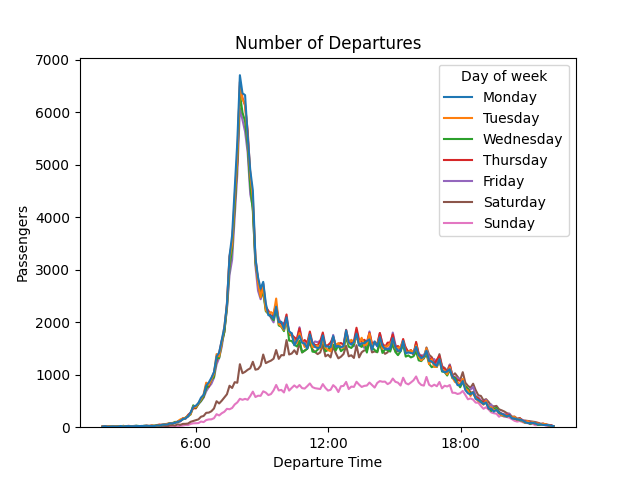}
	\caption{Travel demand over time for days of the week (grouped by starting time)}
	\label{fig:experimental-design:passenger-demand}
\end{figure}

To analyze allocation decisions and the impact of mitigation measures applied to public transport on the system's reliability, we use passenger data generated by the travel demand modeling MITO7 (see Section XY). Figure~\ref{fig:experimental-design:passenger-demand} shows a typical week of travel demand as identified by MITO7. We observe a spike in demand in the morning on weekdays and the expected low demand in the early morning and late afternoon. Note that the grouping only reflects the starting time of passenger demand and not how many passengers may use the system concurrently throughout their travel. 

We focus our studies on the time frame between 6 a.m. and 12 p.m., which includes the highest demand. Analog to the Simulation approach, we consider a 25\% subset of the travel demand. We begin our analysis by evaluating the status quo, i.e., how the system performs without any restrictions. Our base scenario includes all passenger demand for the selected time frame from the 25\% subset and allocates one vehicle to every transit route. Note that we scale the capacity of an allocation, i.e., a vehicle, to 25\% of its actual capacity to match the subset size of the passenger demand.

To investigate the effects of capacity restrictions reflecting mitigation strategies directed either by governmental decree or by the transport service provider herself, we consider three settings: i) full capacity, ii) half capacity, and iii) one-third capacity. Furthermore, we assume decreasing passenger demand depends on the epidemic situation and policies enacted. Herein, we consider several degrees of demand reductions in a sensitivity analysis to identify settings where the transport network will experience bottlenecks and subsequent reduced operational performance, i.e., demand served. At this point, we will draw parallels to the policy impact analysis and simulation findings to identify plausible ranges of settings and their evaluated impacts on the system.

Next, we will consider larger fleet sizes and possibly allocating additional vehicles to bolster routes. We assume that one additional vehicle might be allocatable to any transit route without significantly impacting the schedule. As an alternative, we will consider a scenario where transit routes might not be served, but vehicles will be reallocated to different routes of the same mode.

Finally, we will conclude our experiments with scenarios considering a reduced fleet, where, for example, a sizeable portion of the public transit staff might not be available due to infections.

The experiments were conducted on the same cluster setup as described before. However, we limit the runtime of one scenario to 2 hours.

\paragraph{Results.}
We begin our analysis with the status quo, i.e., how the system currently performs and how the current allocations would fare with decreasing capacity and passenger demand. Figure~\ref{fig:results:status-quo} shows the demand served (\ref{fig:results:status-quo:demand-served}) and the average passenger travel time (\ref{fig:results:status-quo:time}) for different demand levels, ranging from all demand (100\%) to 10\% in 10\% decrements. Note that we consider 100\% to be the full demand of the 25\% subset we described in Section~\ref{section:experimental_design}, which implies that 10\% is 2.5\% of the total demand of the day.

We observe that the system operates on an almost 100\% level of demand served for all passenger demand if the full capacity of the vehicles is available (Note that due to inconsistencies in the data, some demands may not be servable due to spatial or time-related constraints; we did not preprocess the data to remove those passengers and reported the results for the complete subset). 
However, with a decreasing demand, we see a first decline in 40-50\% of the passenger demand, with a final demand served of around 98\%. For the more restricted case with 1/3 capacity, we observe an earlier decline at 20-30\% with a more rapid decrease down to 94\% of demand served. 
When we look at the impact on the average travel time of served passengers, we can see no significant effect on the average time, with average travel times ranging between 2 minutes. 

\begin{figure}[!htb]
	\centering
	\begin{subfigure}[b]{0.48\textwidth}
		\centering
		\newcommand{\thissubplotwidth}[0]{1\textwidth}
\newcommand{\thissubplotheight}[0]{6cm}


		\caption{Passenger demand served.}
		\label{fig:results:status-quo:demand-served}
	\end{subfigure}
	\hfill
	\begin{subfigure}[b]{0.48\textwidth}
		\centering
		\newcommand{\thissubplotwidth}[0]{1\textwidth}
\newcommand{\thissubplotheight}[0]{6cm}

%
		\caption{Average passenger travel time.}
		\label{fig:results:status-quo:time}
	\end{subfigure}
	\caption{Impact of capacity restrictions on the status quo for different demand settings.}
	\label{fig:results:status-quo}
\end{figure}

Figure~\ref{fig:increasing_fleet} shows the results for increasing fleet sizes, i.e., an increase in the number of available allocations per mode by either 10\% or 20\%. Each graph shows the served demand in percent for passenger demand scenario starting at 40\% up to 100\% for a fixed capacity setting. As expected, we can observe that in a full-capacity setting, an increase in fleet size does not improve the performance of the system significantly. However, with 1/2 capacity available, the demand served can be kept at a high level for up to 60\% for a 10\% increase in fleet size and up to 80\% of passenger demand with an increase of 20\% of the fleet. For the lower capacity of 1/3, we see that the demand cannot be served at a high level even for 40\% of the passenger demand; however, with an increase of 20\% in fleet size, 2\% of demand can be additionally served in the full passenger demand setting. 
Note that our approach struggles to find good quality solutions within the runtime of 2 hours for the 10\% increased fleet size and 90-100\% passenger demand. 
In these cases, we note that the solutions of the base case (100\% fleet) are valid solutions for the increased fleet scenarios. However, we can not derive any meaningful insights into whether the results of larger fleets may lead to strictly better performance than the base case.

\begin{figure}
	\newcommand{\thissubplotwidth}[0]{0.33\textwidth}
\newcommand{\thissubplotheight}[0]{6cm}


	\caption{Served demand with increasing budget for different countermeasure and demand scenarios.}
	\label{fig:increasing_fleet}
\end{figure}

\begin{figure}
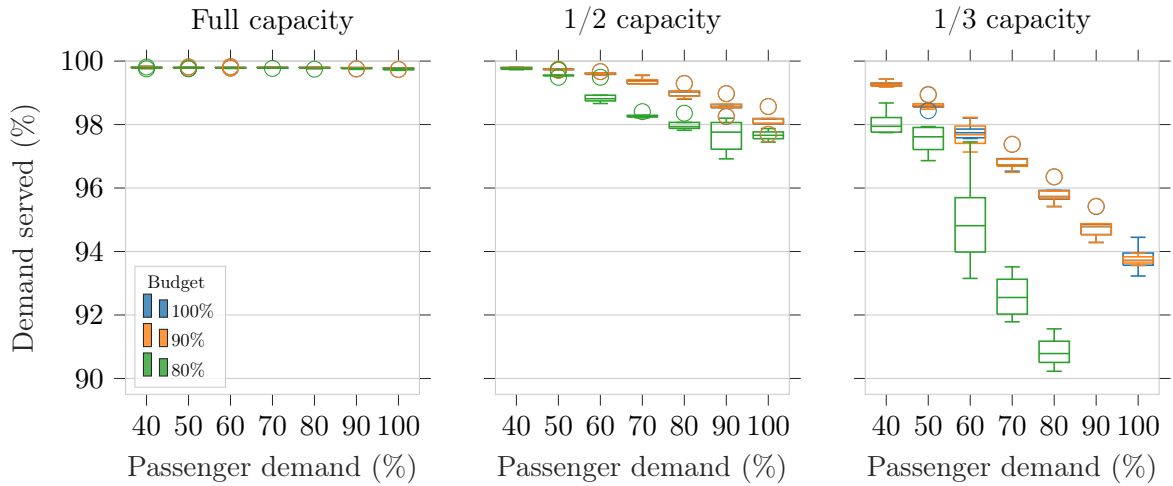

	\newcommand{\thissubplotwidth}[0]{0.33\textwidth}
\newcommand{\thissubplotheight}[0]{6cm}


	\caption{Served demand with decreasing budget for different countermeasure and demand scenarios.}
	\label{fig:decreasing_fleet}
\end{figure}

We now continue with the case where the available budget regarding allocations, i.e., available fleet, is decreasing, shown in Figure~\ref{fig:decreasing_fleet}. A first insight can be drawn from the full-capacity results: even a decrease in budget by 20\% does not significantly deteriorate the demand served by the system. 
This effect is not mirrored for the 1/2 and 1/3 capacity scenarios, where a small deterioration of demand served can be observed when decreasing the budget by 20\% for the 1/2 capacity case and a major decline in the 1/3 capacity scenarios, starting from 60\% passenger demand.
Of note is the performance of the 90\% budget scenarios: we observe that the system's performance regarding demand served is almost equal even when considering a reduced budget.

\begin{figure}[!htb]
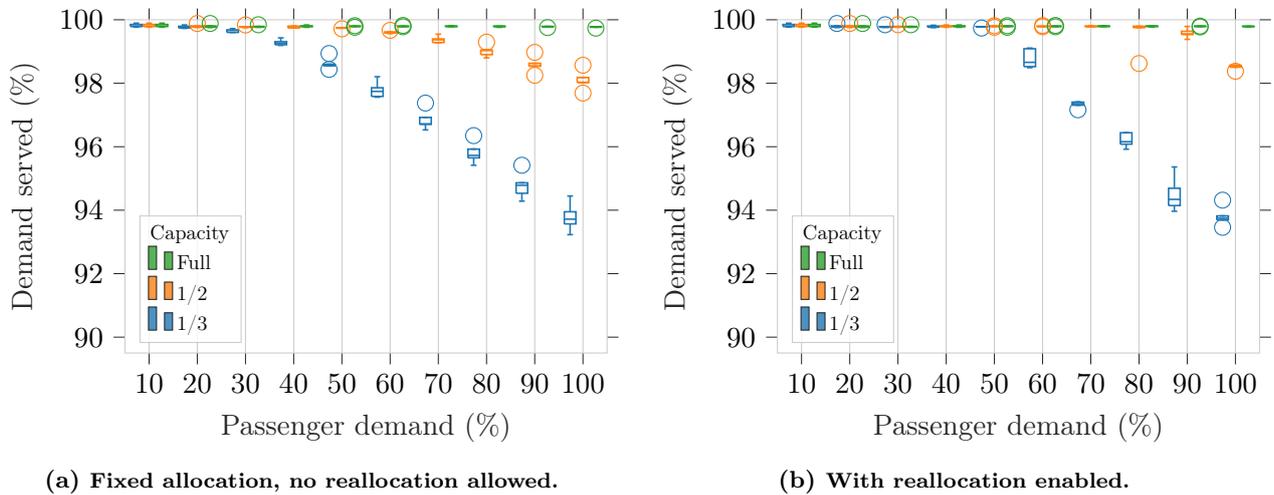

	\centering
	\begin{subfigure}[b]{0.48\textwidth}
		\centering
		\newcommand{\thissubplotwidth}[0]{1\textwidth}
\newcommand{\thissubplotheight}[0]{6cm}


		\caption{Fixed allocation, no reallocation allowed.}
		\label{fig:results:study:computational-study-unassigned-per-fleet}
	\end{subfigure}
	\hfill
	\begin{subfigure}[b]{0.48\textwidth}
		\centering
		\newcommand{\thissubplotwidth}[0]{1\textwidth}
\newcommand{\thissubplotheight}[0]{6cm}

%
		\caption{With reallocation enabled.}
		\label{fig:results:study:computational-study-distance-per-fleet}
	\end{subfigure}
	\caption{Impact of reallocation on the network performance.}
	\label{fig:results:study:computational-study-per-fleet}
\end{figure}

Our last case investigates the effect of allowing re-allocations of existing transit routes to bolster other routes during the considered time frame. The results presented in Figure~\ref{fig:results:study:computational-study-per-fleet} show that even for 1/3 of capacity, the existing routes can serve up to 50\% of the passenger demand, in comparison with 30\% without reallocations. Similarly, for the 1/2 capacity case, the passenger demand served at a high level is increased by 20\% as well (without reallocations: up to 60\%; with reallocation: 80-90\%). 

\section{Travel demand generation module details}
\label{appendix:mito}

In this section we elaborate on the four major modules of the travel demand generation model MITO. The modules are trip generation, destination choice, mode choice and preferred arrival time.

\begin{enumerate}
    \item Trip generation: The number of trips made by an individual is determined using a hurdle model. This approach begins by estimating the probability that a person makes no trips, using a binary logistic regression. If at least one trip is made, a truncated negative binomial regression is then applied to estimate the total number of trips.
    \item Destination choice: The locations for work and school are pre-assigned in the synthetic population. All other types of destinations are determined using a destination choice model based on the logit formulation, which takes into account both travel distance and the attractiveness of each destination. The utility for each potential origin-destination pair, specific to each trip purpose, is defined as follows
    \begin{equation*}
        e^{U_{i|j}} = e^{\beta^* \cdot \text{imp}_{i|j} + \ln(\text{attraction})} = e^{\beta^* \cdot \text{imp}_{i|j}} \cdot \text{attraction},
    \end{equation*}
    In this expression, $e^{U_{i|j}}$ represents the exponential of the utility associated with choosing destination $j$ from origin $i$ for a given trip purpose $p$, while $\text{imp}_{i|j|p}$ denotes the impedance for that trip. The attraction term is estimated during the trip generation step and reflects the number of opportunities available in the destination zone relevant to the trip purpose. The strength of the impedance effect is controlled by the purpose-specific parameter $\beta$. Impedance is calculated as follows:
    \begin{equation*}
        \text{imp}_{i|j} = e^{\text{td}_{i|j}} \cdot c_p,
    \end{equation*}
    Here, $\text{td}_{i|j}$ represents the travel distance between origin $i$ and destination $j$, and $c_p$ is a calibrated constant specific to each trip purpose $p$. Travel distance serves as the basis for calculating impedance. The parameters $\beta$ and $c_p$ are adjusted during calibration to reflect both the distribution and the average trip lengths reported for each purpose in the German household travel survey.
    \item Mode choice: 
    Mode choice is determined using a nested logit discrete choice model, which selects the travel mode for each trip based on a combination of trip attributes, mode-specific characteristics, and individual traveler features. The available travel modes include Auto driver, Auto passenger, Bicycle, Bus, Train, Tram/Metro, and Walk. These are grouped into nests, with Auto driver and Auto passenger forming the Auto nest, and Bus, Train, and Tram/Metro forming the Transit nest. The probability of selecting a specific mode is given by:

    \begin{equation*}
        \text{Pr}(i) = \frac{e^{V_i}}{\sum_{j=1}^{J} e^{V_j}} \quad 
    \end{equation*}
with $\text{Pr}(i)$ denoting the probability of choosing alternative $i$, and $V_j$ as the observable component of utility of alternative $j$. 
In the case of nested modes, this probability is computed as the product of the probability of choosing the nest and the conditional probability of selecting a specific mode within that nest.

The systematic (observable) utility component for an individual $t$ choosing alternative $i$ is defined as:

\begin{equation*}
V_{i,t} = V(S_t) + V(X_i) + V(S_t, X_i)
\end{equation*}

Here, $V(S_t)$ represents the utility derived from the individual’s characteristics, $V(X_i)$ captures the contribution from mode-specific attributes, and $V(S_t, X_i)$ accounts for interaction effects between the individual's characteristics and the attributes of the travel mode.
    \item Preferred arrival time: Work and school purpose trips have arrival time provided in the synthetic population.For all other purpose trips, the preferred departure time is given probabilistically from observed arrival time distributions.
\end{enumerate}

\end{appendices}

\end{document}